\documentclass[useAMS,usenatbib,twocolumn,fleqn]{mnras}
\usepackage[T1]{fontenc}
\usepackage{ae,aecompl}
\usepackage{natbib, graphics, epsfig}
\usepackage{stmaryrd}

\usepackage{bm}
\usepackage{times}
\usepackage{amsmath}
\usepackage{amssymb}
\usepackage{textcomp}
\usepackage{verbatim}

\newcommand{\msun}{{\,\rm M_\odot}}
\newcommand{\kms}{\,{\rm km}\,{\rm s}^{-1}}

\newcommand{\erg}{\,{\rm erg}}
\newcommand{\Gyr}{\,{\rm Gyr}}

\newcommand{\pc}{\,{\rm pc}}
\newcommand{\kpc}{\,{\rm kpc}}
\newcommand{\Mpc}{\,{\rm Mpc}}

\newcommand{\cpm}{\,{\rm cm}^2\,{\rm g}^{-1}}
\renewcommand{\v}[1]{\ensuremath{\mathbf{#1}}} % for vectors

\let\baraccent=\= % rename builtin command \= to \baraccent
\renewcommand{\=}[1]{\stackrel{#1}{=}} % for putting numbers above =

\newcommand{\IChalf}{\chi^2_{\rm init}=50\%}
\newcommand{\ICfull}{\chi^2_{\rm init}=100\%}

\usepackage{color}

\def\jcap{J. Cosmol.  Astropart. Phys.}
\def\aap{A\&A}
\def\apj{ApJ}

\def\apjl{ApJ}
\def\mnras{MNRAS}
\def\araa{ARA\&A}

\def\nat{Nature}

\def\prd{Phys. Rev. D}
\title[Evaporating the Milky Way with inelastic SIDM]
      {Evaporating the Milky Way halo and its satellites with inelastic self-interacting dark matter}
      \author[M. Vogelsberger et al.] {\parbox{18.5cm}{
	  Mark Vogelsberger$^1$\thanks{e-mail: mvogelsb@mit.edu}\thanks{Alfred P. Sloan Fellow}, 
          Jes\'us Zavala$^2$,
          Katelin Schutz$^{3}$
          and Tracy R. Slatyer$^{1,4}$ 
        }\vspace{0.3cm}\\
        $^1$ Department of Physics, Kavli Institute for Astrophysics and Space Research, Massachusetts Institute of Technology, Cambridge, MA 02139, USA\\	
        $^2$ Center for Astrophysics and Cosmology, Science Institute, University of Iceland, Dunhagi 5, 107 Reykjavik, Iceland \\
        $^3$ Berkeley Center for Theoretical Physics, University of California, Berkeley, CA 94720 \\
        $^4$ Center for Theoretical Physics, Massachusetts Institute of Technology, Cambridge, MA 02139, USA}

\date{Accepted XXX. Received YYY; in original form ZZZ}

\pubyear{2018}

\begin{document}
\label{firstpage}
\pagerange{\pageref{firstpage}--\pageref{lastpage}}
\maketitle

%\begin{document}
%\date{Accepted ???. Received ???; in original form ???}

%\pagerange{\pageref{firstpage}--\pageref{lastpage}} \pubyear{2015}

%\maketitle

%\label{firstpage}

\begin{abstract}
Self-interacting dark matter provides a promising alternative for the cold
dark matter paradigm to solve potential small-scale galaxy formation
problems. Nearly all self-interacting dark matter simulations so far
have considered only elastic collisions. Here we present simulations of a galactic halo within a generic inelastic model using a novel
numerical implementation in the {\sc Arepo} code to study arbitrary
multi-state inelastic dark matter scenarios. For this model we find that inelastic
self-interactions can: (i)~create larger subhalo density cores compared to elastic
models for the same cross section normalisation; (ii)~lower the abundance of
satellites without the need for a power spectrum cutoff;
(iii)~reduce the total halo mass by about $10\%$; (iv)~inject the energy
equivalent of $\mathcal{O}(100)$ million Type II supernovae in galactic haloes through level de-excitation; (v)~avoid the gravothermal catastrophe due
to removal of particles from halo centres. We conclude that a $\sim
5$ times larger elastic cross section is required to achieve the same
central density reduction as the inelastic model. This implies that well-established constraints on self-interacting cross sections have
to be revised if inelastic collisions are the dominant mode. In this case
significantly smaller cross sections can achieve the same core density reduction thereby increasing the parameter space of allowed models considerably.
\end{abstract}


\section{Introduction}

The Cold Dark Matter (CDM) paradigm has been proven to be very successful in describing
the large-scale distribution of galaxies~\citep[e.g.][]{Springel2005} and serves as the cornerstone of our current
understanding of galaxy formation and evolution~\citep[e.g.][]{Vogelsberger2014a,Schaye2015,Khandai2015,Dubois2016,Springel2018}.  
At sub-galactic scales however, the CDM paradigm remains to be verified with various
outstanding challenges that have appeared over the last decades~\citep[for a recent review see][]{Bullock2017}.
Among the most relevant challenges are: the under-abundance of dwarf galaxies in the Milky Way (the missing satellites problem,~\citealt{Klypin1999,Moore1999}) and in the field~\citep{Zavala2009,Papastergis2011,Klypin2015}, the
unexpected inner dark matter density profile in low surface brightness galaxies and dwarf galaxies (the cusp-core
problem, e.g.~\citealt{deBlok1997,Walker2011}), the deficit of dark matter in the inner regions of massive dwarf galaxies (the
too-big-to-fail problem, e.g.~\citealt{BK2011,Papastergis2015}), and the unexpected diversity in the shape of dwarf rotation curves~\citep{Oman2015,Oman2016}.

Most of these problems have been identified by contrasting dark-matter-only
simulations with observations, which is clearly an oversimplified comparison that does not take into account the complex baryonic physics at play. It is
therefore possible that some or even all of these challenges to CDM 
can be solved through the proper modelling of baryonic physics and by
carefully considering observational biases. Plausible solutions have been presented
separately for each problem. For instance, the existence of dark matter cores
could be explained by the gravitational transfer of energy from supernovae into the orbits
of dark matter particles~\citep[e.g.][]{Navarro1996,Governato2012,Onorbe2015,Chan2015, Read2016}.
 The resulting inner dark matter densities, in combination with strong environmental
effects like tidal stripping and heating, have been invoked to alleviate the too-big-to-fail problem in
the Milky Way~\citep[e.g.][]{Zolotov2012,Brook2015,Sawala2016,Wetzel2016}. Furthermore, accounting for observational biases influences the interpretation of the severity of the
dwarf abundance problem (in the Milky Way satellites e.g.~\citealt{Koposov2008,Kim2017} and in the field e.g.~\citealt{Brooks2017}), 
as well as the too-big-to-fail problem in the field~\citep[e.g.][]{Dutton2016,Verbeke2017}, and more recently the diversity
problem of dwarf rotation curves~\citep{Santos2018}.

Nevertheless, to this date, a comprehensive solution to all these CDM challenges remains elusive since
there is yet no consensus on a definitive implementation of baryonic physics galaxy formation models. In particular, it is uncertain how to couple the sub-resolution physics of the supernova
explosion with the effective energy injection into the interstellar medium, and ultimately into the dark matter
distribution. This coupling between supernovae and dark matter, which is seemingly crucial in solving the CDM challenges, 
depends on the stellar mass content of the galaxy relative to
the depth of the potential well~\citep[e.g.][]{Penarrubia2012,DiCintio2014}, which sets the energy requirements for
the cusp-core transformation: the smaller the galaxy, the less likely it is that this transformation is viable.
Another key ingredient is how efficiently the available energy is deposited in the interstellar medium, and on which timescales it
is injected relative to the dynamical time scales of the inner regions of the halo. Large modifications to the dark matter distribution are associated with
``bursty'' star formation histories, which have been shown to prevail in galaxies with stellar masses in the range
$10^8-10^{10}\msun$~\citep{Kauffmann2014}, however, at lower masses, where the CDM challenges are more 
severe, the time resolution needed to settle this issue remains unaccessible~\citep{Weisz2014}.

Given these outstanding CDM challenges, it is sensible to consider the
possibility that these problems actually have a root in the
the CDM assumptions about the underlying nature of dark matter. The fact that two of the most well-studied CDM candidates, the axion and the thermal WIMP (Weakly Interacting Massive Particle),
have not been detected yet, gives further motivation to study CDM alternatives.
In the context of structure formation theory, the two key assumptions of CDM where 
dark matter is assumed to be cold and collisionless, can be relaxed in the following two ways:
(i) introduction of a galactic-scale cutoff in the linear power spectrum in the early
Universe, either through free streaming of dark matter particles, known as Warm Dark Matter, 
(WDM, e.g.~\citealt{Colin2000,Bode2001}), or through interactions between dark matter
particles and relativistic particles (known as interacting dark matter, e.g.~\citealt{Boehm2002,Buckley2014});
(ii) allowing for strong dark matter self-interactions in the late Universe, known as 
Self-Interacting Dark Matter (SIDM, e.g.~\citealt{Spergel2000}). We note that these two modifications
can actually be present within the same particle model~\citep[e.g.][]{vA2012}, and can
be studied generically within the framework of a generalised theory of structure formation~\citep[see the ETHOS framework, ][]{ETHOS1,ETHOS2,Lovell2018}.

The CDM challenges can be alleviated to different degrees by these two modifications. For
instance, a cutoff in the power spectrum reduces both the abundance of dark matter
structures, and their inner densities, which helps in solving the missing dwarfs challenge~\citep[e.g.][]{Zavala2009} and the too-big-to-fail problem~\citep[e.g.][]{Lovell2012}. In the
context of thermal WDM, the relevance of this possibility as a solution is 
small due to constraints on thermal WDM particles from Ly$-\alpha$
forest observations~\citep[e.g.][]{Viel2013,Irsic2017}, which require $m_{\rm WDM}\gtrsim3.5\,{\rm keV}c^{-2}$,
which is too large to have a meaningful impact on the CDM challenges~\citep[e.g. $m_{\rm WDM}\sim2\,{\rm keV}c^{-2}$][]{Schneider2014},
although see~\citet{Garzilli2017}. In the context of interacting dark matter, dark-matter-only simulations have shown
that this scenario substantially alleviates the missing satellite problem~\citep[][]{Boehm2014}, the too-big-to-fail problem~\citep{ETHOS2}, and the diversity of rotation curves problem for the smallest
dwarfs~(\citealt{ETHOS2}, Moseley et al. 2018 in prep.). Since this model includes 
a complex cutoff in the power spectrum with dark acoustic oscillations, a
detailed analysis with current Ly$-\alpha$ constraints remains to be done in order
to assess how relevant these models remain for the CDM challenges (Bose et al. 2018 in prep.).

Without SN feedback, neither WDM nor interacting DM can result in the creation
of dark matter cores. The inner density profiles have a lower normalisation compared to CDM, but
the profile remains cuspy since there is no relevant mechanism to form a core (primordial thermal
motions can set a maximum to the phase space density, hence a core, but the size of allowed
cores is too small to be astrophysically significant, e.g.~\citealt{Maccio2012, Shao2013}). SIDM is thus
needed as an alternative mechanism to create dark matter cores, by transferring energy from
the outside in, thermalising the inner dark matter distribution~\citep[e.g.][]{Colin2002}. By now, this is 
a well understood process that has been shown to create $\mathcal{O}({\rm kpc})$ size cores in the
centre of haloes with allowed transfer cross sections per unit mass $\sigma_T/m_\chi\sim1\cpm$.
This is enough to alleviate the core-cusp problem and the too-big-to-fail problem~\citep{Vogelsberger2012,Rocha2013,Zavala2013},
and it seemingly enhances the diversity of rotation curve shapes relative to CDM for a fixed
baryonic physics implementation~\citep{Kamada2017,Creasey2017}. Current constraints on the
cross sections are however too strong for SIDM to have a relevant impact on the dwarf
abundance problem (cross sections an order of magnitude larger than current limits are 
needed to evaporate substructure during halo-subhalo particle interactions). 
By now, the SIDM model has been studied extensively both within particle
theories and within structure formation using numerical simulations, becoming
an appealing alternative to the CDM model~\citep[for a recent SIDM review see][]{Tulin2017}.

With a few exceptions~\citep[][]{TM2017a,TM2017b}, structure formation within the SIDM scenario has been restricted to elastic scattering only, i.e., during a collision between dark matter particles, kinetic
energy is conserved. This does not have to be the case however, since well-motivated 
particle models exist that include ``excited'' dark matter states~\citep[e.g.][]{AH2009,Loeb2011,Schutz2015}.
In the simplest two-state case, there is 
a ground ($\chi^1$) and an excited state ($\chi^2$): a transition from the
lower to the upper level, up-scattering, absorbs energy (endothermic reaction), while
down-scattering releases energy (exothermic reaction). The latter
case is especially interesting for structure formation since the preferential velocity kick imparted to the dark matter particles
upon down-scattering from the centre of haloes diminishes the inner dark matter content, augmenting the efficiency of core creation
relative to purely elastic SIDM. In addition, and contrary to the purely elastic case, interactions between halo and subhalo particles 
along the orbit of a subhalo, might lead to velocity kicks that are large enough to evaporate the subhalo, which
might be a viable mechanism to reduce the abundance of dwarf galaxies. Similar velocity kicks have been studied in the case of late decays from an excited-state population, e.g. \cite{Wang2014}, but unlike the decay rate, the scattering rate is velocity-dependent and enhanced in regions of high density.
The magnitude of the velocity kick in the exothermic case depends on the exact
energy splitting, $\delta$, between the ground state, with mass $m_{\chi^1}$ and the excited state, with mass $m_{\chi^2}$,
$v_{\rm kick} = \sqrt{2 \delta/m_{\chi^1}}$. The velocity kick would thus have a substantial impact on the orbits of dark matter 
particles if the mass splitting is comparable to their orbital kinetic energy. Following~\citet{Schutz2015}, if we take
the typical velocity in dwarf galaxies to be ${\sim 10\kms \sim 3
\times 10^{-5}\,c}$, this then implies ${\delta/m_{\chi^1} > 10^{-9}c^2}$ in order for
down-scattering to have a relevant effect. As $\delta \rightarrow 0$,  the model
reduces to the standard elastic SIDM case. Inelastic SIDM models therefore have the potential
to substantially change the population of subhaloes in Milky Way-like haloes. However, this scenario
has so far not been explored. 

\begin{figure}
\centering
\hspace{-0.2cm}\includegraphics[width=0.49\textwidth]{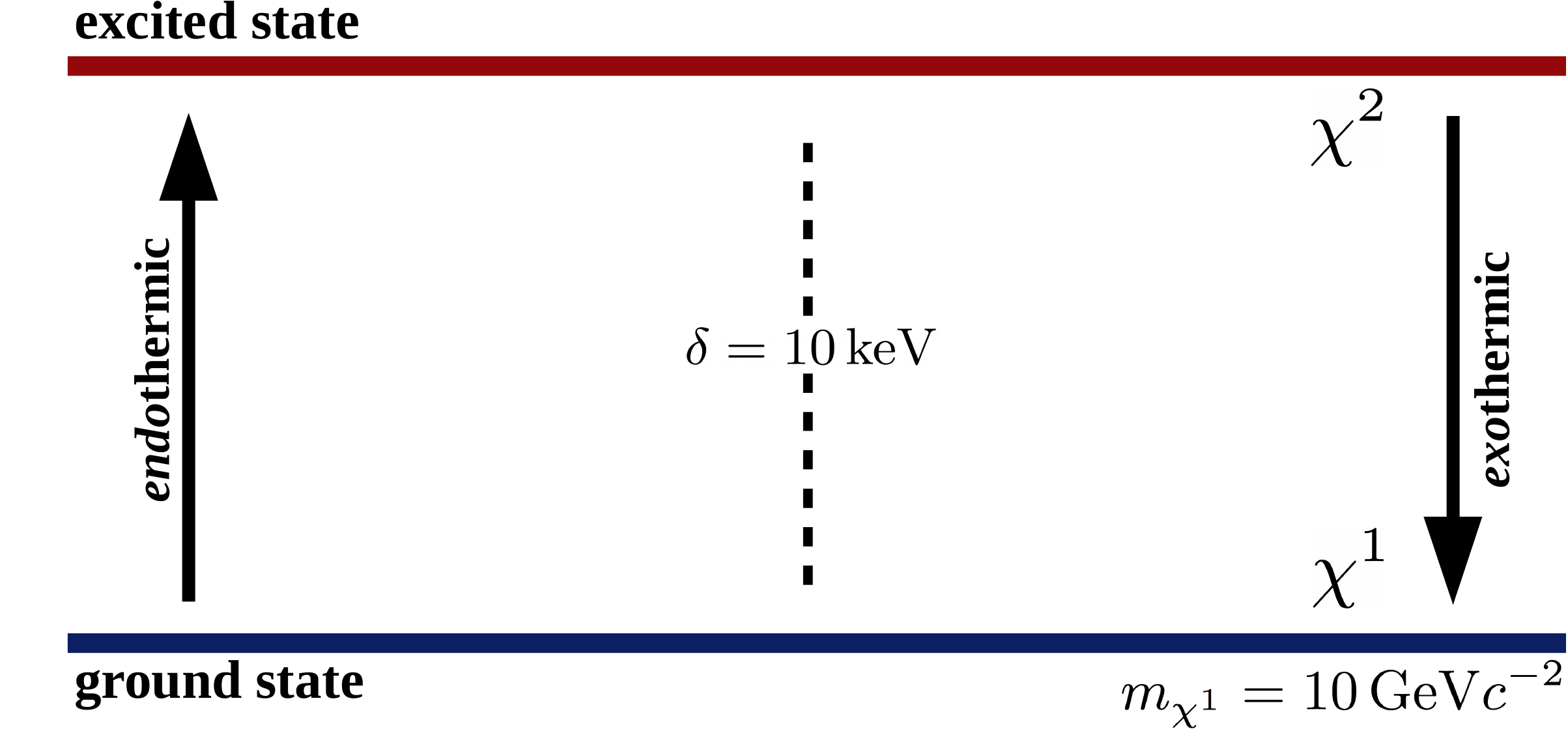}
\hspace{-0.2cm}\includegraphics[width=0.49\textwidth]{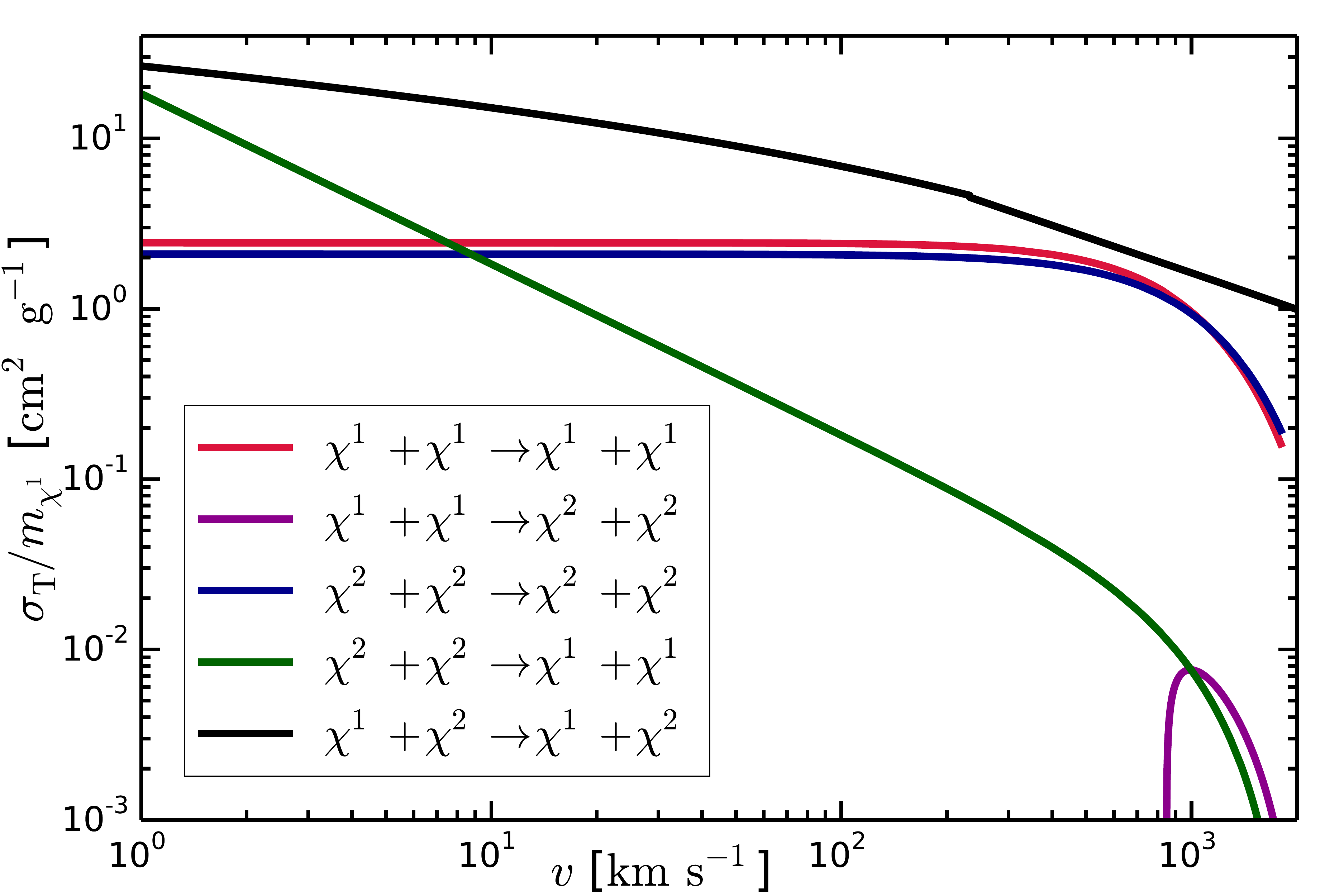}
\caption{{\bf Inelastic self-interacting dark matter model.} {\it Top panel:} Schematic overview of the two-state inelastic SIDM model. The
ground state ($\chi^1$) and excited state ($\chi^2$) are split by $\delta =
10\,{\rm keV}$. This two-state model allows for exo- and endothermic reactions. {\it Bottom panel:}  Self-scattering cross section per unit mass for the different reactions of the
two-state inelastic dark matter model. Up-scattering (${\chi^1 + \chi^1 \rightarrow \chi^2 +
\chi^2}$, purple) is suppressed in the Milky Way environment since it can only occur for large relative velocities,
${v > 2 \sqrt{2 \delta/m_\chi^1} \cong 848\kms}$. The elastic cross sections (${\chi^1 + \chi^1 \rightarrow
\chi^1 + \chi^1}$, red) and (${\chi^2 + \chi^2 \rightarrow \chi^2 + \chi^2}$, blue) are
nearly flat over the whole velocity range relevant for the Milky Way environment. Down-scattering (${\chi^2 +
\chi^2 \rightarrow \chi^1 + \chi^1}$, green) mainly occurs for rather low velocities ($v \lesssim 10\kms$).
This reaction is exothermic corresponding to a velocity kick of ${\sqrt{2 \delta/m_\chi^1} \cong 424\kms}$.
The largest cross sections occur for mixed-state elastic scattering
(${\chi^1 + \chi^2 \rightarrow \chi^1 + \chi^2}$, black), with a normalisation that  exceeds ${\sim 10\cpm}$ for low
velocities.}
\label{fig:schematic}
\end{figure}

The goal of this paper is to study the impact of inelastic SIDM on galactic
haloes. We extend the algorithm presented in~\citet{Vogelsberger2012} to
include multiple particle species to perform high-resolution
simulations of a galactic halo within an inelastic SIDM model.  The paper is
structured as follows. In Section 2 we present the inelastic SIDM model that we
explore in this paper. In Section 3 we describe
in detail the numerical implementation of the inelastic SIDM algorithm in the simulation
code {\sc Arepo}. In Section 4 we explore the impact of inelastic SIDM on the
dark matter distribution of our simulated Milky Way-sized halo and its
substructure.  Finally, in Section 5 we present our conclusions.

\section{Inelastic SIDM model}

We aim for a numerical exploration of the effects of inelastic dark matter 
self-interaction on the evolution of structure of a galactic dark matter halo. This requires a simple dark matter particle physics
model, which has a tractable amount of free parameters such that numerical
studies can be used to sample the relevant parameter space. Here, we use the
two-state dark matter model of self-scattering including a nearly-degenerate excited
state that has been presented in~\cite{Schutz2015}. This model is based on
accurate analytic approximations for the corresponding elastic and inelastic s-wave cross
sections, which are valid outside the perturbative regime provided the particle
velocity is sufficiently low. 

The underlying particle physics model is characterised by four parameters:
the mass splitting $\delta$ between ground ($\chi^1$) and excited ($\chi^2$) state, the coupling
constant $\alpha$ between the dark matter particle ($\chi^{1,2}$) and the mediator ($\phi$),
the mass of the dark matter particle $m_{\chi^{1,2}}$, and the mass of the force mediator
$m_\phi$. We note that although this is clearly not the only well-motivated dark matter  model with inelastic self-scattering, it is the only such model
where the description of scattering at the velocity scales of dwarf galaxies
has proven analytically tractable thus far. 
Furthermore, the involved Yukawa
potential has proven ubiquitous in different areas in particle physics,
rendering this model a useful representative example for studies of inelastic dark matter self-scattering more generally. 

Small mass splittings between states, similar to those we consider, could also
be present in the context of atomic dark matter~\citep[e.g.][]{Kaplan2010, Fan2013, Foot2015, Choquette2015, Ghalsasi2018}, non-Abelian dark sectors where
the dark matter is part of a nearly-degenerate multiplet of states~\citep[e.g.][]{Chen2009, Cirelli2010}, or in scenarios where the DM forms stable bound
states~\citep[e.g.][]{Wise2014}. However, it should be noted that the scenario we
consider requires that transitions from the heavier to the lighter state occur
primarily through collisional de-excitation; it is necessary but not sufficient
to have small mass splittings. In models that include a very light force
carrier, such as classic dark atom models, the excited states are typically
depleted by radiative decay rather than collisions, leading to different
effects on structure formation~\citep[e.g.][]{Wang2014}. Our calculation has other
qualitative features that are not universal to dark-sector models with small
mass splittings: for example, that there are only two states that significantly
participate in the phenomenology, and that de-excitation requires two
excited-state particles to simultaneously de-excite. Thus while this model
serves as a valuable illustrative example of the effects of inelastic
self-scattering, detailed quantitative results for other dark-sector models
would generally require dedicated analyses.

\cite{Schutz2015} found that at the velocity scale of ${\sim 10\kms}$, 
relevant for the dynamics of dwarf galaxies, the range of
particle physics parameters that result in interesting cross sections 
are: ${m_{\chi^1} \in [0.1, 300]\,{\rm
GeV}c^{-2}}$, ${m_\phi \in [10^{-4}, 1]\,{\rm GeV}c^{-2}}$, ${\alpha \in
[10^{-3},10^{-1}]}$, ${\delta \in [0.1, 10]\,{\rm keV}}$.  We select
one particular model from this range with ${\delta = 10\,{\rm keV}}$,
${m_{\chi^1}=10\,{\rm GeV}c^{-2}}$, ${m_\phi=30\,{\rm MeV}}$ and ${\alpha=0.1}$ (see top panel of Fig.~\ref{fig:schematic} for a schematic representation). This choice results in an elastic cross section per unit mass\footnote{We remark that for the model considered, the
relevant cross section, which is the transfer cross section is the same as the total cross section given the lack of angular dependence
of the differential cross section, see Section B3.3 of~\citet{Schutz2015}. Hereafter we use the terms transfer cross section and cross section interchangeably.} of a few ${\rm cm^2}{\rm g}^{-1}$ at the scale of
dwarf galaxies, while the velocity kick is 
of the order of ${v_{\rm kick}\sim424\kms}$.
We remark that this model has not been fine-tuned, and only represent a benchmark point in the relevant
parameter space. The model has five different reactions and corresponding cross
sections, which are presented in the bottom panel of Fig.~\ref{fig:schematic}: elastic scattering in
the ground state (${\chi^1 + \chi^1 \rightarrow \chi^1 + \chi^1}$), elastic
scattering in the excited state (${\chi^2 + \chi^2 \rightarrow \chi^2 +
\chi^2}$), endothermic up-scattering (${\chi^1 + \chi^1 \rightarrow \chi^2 +
\chi^2}$), exothermic down-scattering (${\chi^2 + \chi^2 \rightarrow \chi^1 +
\chi^1}$), and elastic Yukawa scattering (${\chi^1 + \chi^2 \rightarrow \chi^1
+ \chi^2}$). We note that our benchmark point practically forbids
up-scattering for the typical velocities of dark matter particles in the Milky Way environment, 
since the required energy splitting is too large for this reaction to occur frequently.  
This can be seen in the bottom panel of Fig.~\ref{fig:schematic}, where the cross section for up-scattering is zero for relative velocities ${v < 2\sqrt{2 \delta/m_{\chi^1}} \cong 848\kms}$. 

Besides specifying the model parameters, we also have to specify the initial
conditions for the abundance of dark matter species; i.e. what fraction of dark matter is in which state initially -- ground
state ($\chi^1$) or excited state ($\chi^2$).  Obviously, putting all particles
initially into the excited state will maximise the effect of energy release
during structure formation.  On the other hand, putting all particles initially
in the ground state will do the opposite, and behave like a purely elastic
SIDM model with multiple different cross sections. To explore the relevant range for
inelastic SIDM, we consider in the following two initial configurations: all
particles initially, at $z=127$ as described below, in the excited state ($\ICfull$), or half the particles initially in the excited state
($\IChalf$). 

We note that in the generic framework of the model we are considering here,
the presence of a potential mediated by light vector exchange automatically
implies that both the ground and excited states can annihilate into the light
vectors. For the large dark-sector coupling ($\alpha = 0.1$) and relatively
light DM mass ($10\,{\rm GeV}c^{-2}$) considered in this work, the natural cross section for
this process is ${\langle \sigma v \rangle = \pi \alpha^2/m_\chi^2 \sim 3
\times 10^{-21}\,{\rm cm^3}{\rm s}^{-1}}$, which is much larger than the thermal relic cross
section, implying a depletion of the dark matter states in the early Universe.
To justify a large primordial abundance of the excited state, a non-thermal
origin for these species could be invoked, which could occur through the decay
of thermally produced heavier species, preferentially into the excited state
(as discussed in \citealt{Loeb2011}). In order to have $\sim100\%$ of all
particles in the excited state by the starting redshift of the simulations
$z=127$, it is necessary to suppress down-scattering (or delay the production of the excited state) until around the epoch of recombination ($z=1100$).

If instead, down-scattering is allowed already by the time of matter-radiation
equality ($z_{\rm eq}\sim3400$), then we can make a simple estimate of the ratio of 
excited to ground states by the time the simulation starts at $z_{in}=127$. 
\begin{equation}
\chi=1-\int_{z_{\rm in}}^{z_{\rm eq}}\frac{\Gamma(z)}{H(z)(1+z)}dz
\end{equation}
where $H(z)$ is the Hubble expanstion rate and $\Gamma(z)$ is the scattering rate for
de-excitation:
\begin{equation}
\Gamma(z)=\rho_{\rm dm}(z)v_{\rm typ}(z)\sigma^{\rm de}_T(v_{\rm typ}(z))/m_\chi
\end{equation}
where $v_{\rm typ}$ is the characteristic velocity of dark matter particles, which at
redshifts prior to $z=127$ is roughly equal to the velocity dispersion of
unclustered dark matter particles 
$v_{\rm typ}\sim T(z)/(m_\chi T_{\rm kd})^{1/2}$
where $T(z)$ is the radiation temperature and $T_{\rm kd}$ is the dark matter kinetic 
decoupling temperature. After decoupling, the dark matter temperature and radiation temperature scale as
$T_\chi=T^2/T_{\rm kd}$. We fix $T_{\rm kd}=10$~MeV\footnote{In Eq. 25 of \citet
{Feng2010}, this choice of $T_{\rm kd}$ would
correspond to a kinetic mixing parameter $\epsilon\sim3\times10^{-7}$.}, but note that the dependence of $\chi$ on $T_{\rm kd}$ is
only mild. The cross section for de-excitation
$\sigma^{\rm de}_T/m_\chi$ is extrapolated down to the very small typical velocities
of dark matter particles at early redshifts, from the behaviour in the
bottom panel of Fig.~\ref{fig:schematic}. This calculation results in
$\chi\sim0.76$.

If the excited state is populated and down-scattering is unsuppressed at times prior to matter-radiation equality, but after the dark matter temperature drops below the mass splitting, then down-scattering could deplete the excited-state population by a larger factor \citep[][]{Loeb2011}. 

Plausible models with late-time decays would need to avoid a series of
constraints in the early Universe, for instance in their gravitational effects
on the cosmic microwave background radiation~\citep[e.g.,][]{Poulin2016} and large scale structure. Such constraints may point toward models where the metastable decaying species is only slightly heavier than the dark matter. Since
our focus in this paper is on the implications for structure formation of a
large metastable population of excited states in the early Universe, we leave a
detailed study of possible models for later work. Models with large late-time populations of a metastable excited state have been previously discussed in the context of indirect-detection and direct-detection signals~\citep[e.g.][]{Finkbeiner2009, Cline2011}.

\section{Numerical Implementation in {\sc Arepo}}
\label{sec:numerics}

Although we study in this paper a specific two-state inelastic SIDM model, we have implemented a
more general multi-state dark matter framework into the {\sc Arepo}
code~\citep[][]{Springel2010}.  This framework is able to handle an arbitrary
number of states with an arbitrary number of reactions and corresponding cross
sections, and with arbitrary, also non-degenerate, energy level splittings. This code represents a generalisation and 
complete rewrite of the algorithm presented in~\cite{Vogelsberger2012} and \cite{ETHOS2}, which has been employed in multiple
previous SIDM studies~\citep[][]{Vogelsberger2013a, Vogelsberger2014c, Vegetti2014, Dooley2016, Brinckmann2018, Sameie2018}. Here we briefly describe this new numerical implementation.

\begin{figure*}
\centering
\includegraphics[width=0.48\textwidth]{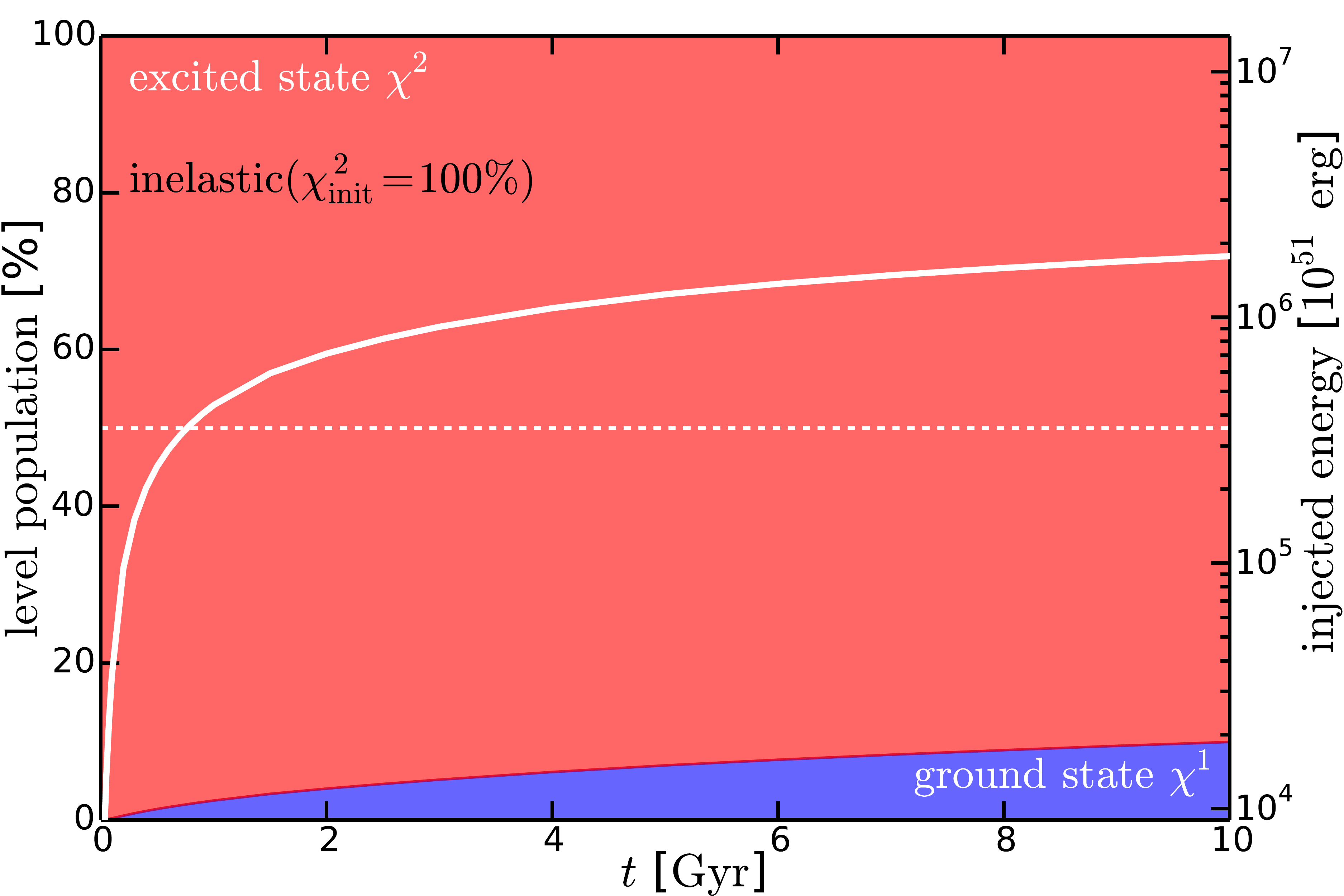}
\hspace{0.5cm}\includegraphics[width=0.48\textwidth]{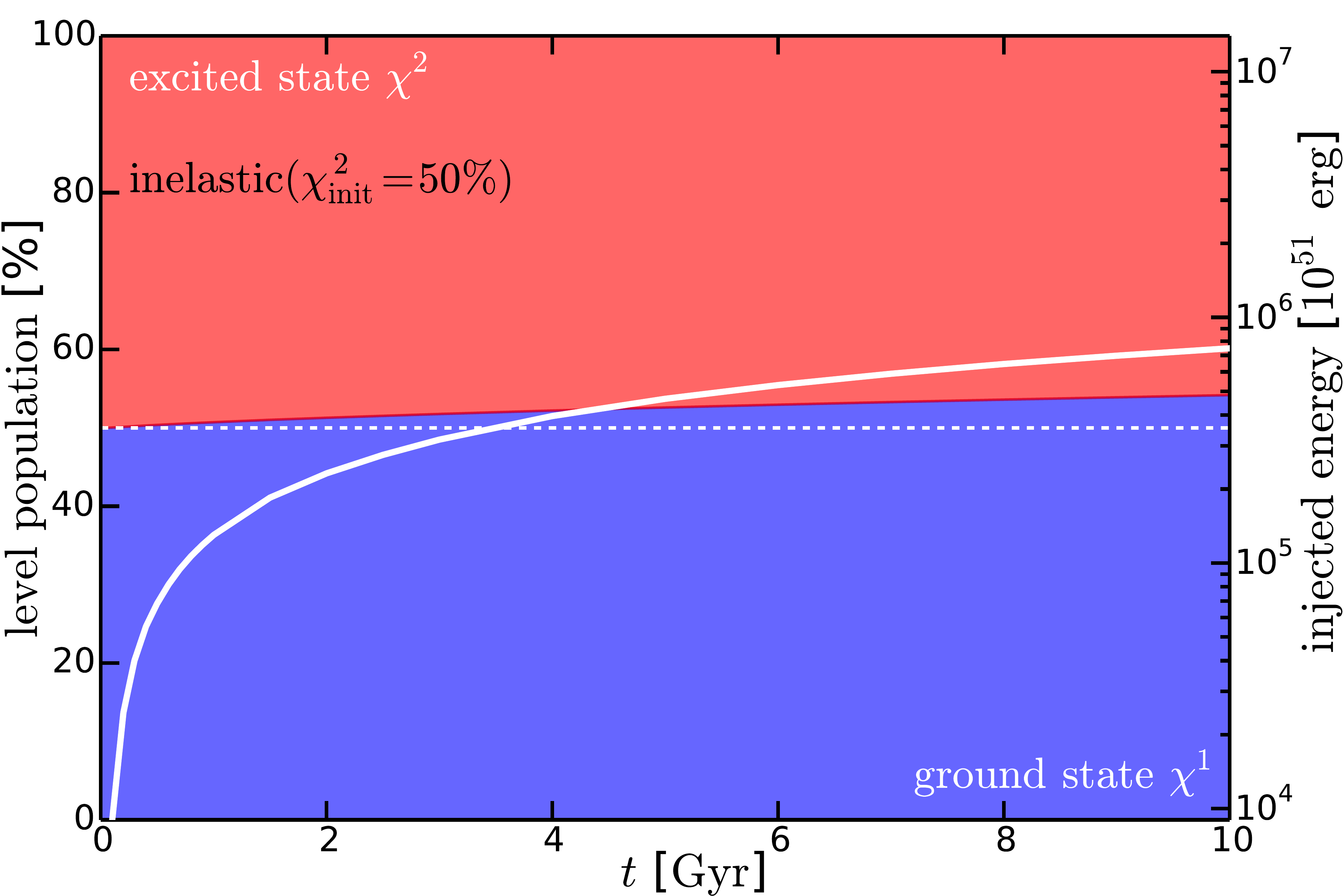}
\caption{{\bf Time evolution of the level population for the inelastic SIDM model with different initial population split.} For the $\ICfull$ configuration ({\it left panel}) all particles are initially in the excited
state ($\chi^2$), whereas for the $\IChalf$ configuration ({\it right panel}) only
$50\%$ of the particles are initially excited and the rest is in the ground
state ($\chi^1$). The split between the two level populations is shown by the filled
areas. The solid white lines present the total energy injected into the halo
through level decay.  This energy is plotted in units of the canonical energy of a
single SNII ($10^{51}\erg$) with the scale shown on the right vertical axis. After $10\Gyr$, de-excitation has injected
$\sim 2\times 10^{57}\erg$ ($\sim 7 \times 10^{56}\erg$) for the $\ICfull$ ($\IChalf$) configuration.}
\label{fig:level_evolution}
\end{figure*}

In the following we assume that each dark matter simulation particle $i$ is in a
specific state $\alpha$, i.e. every simulation particle represents a single state
and not a mixture of different states.  The simulation volume is then filled
with dark matter simulation particles in different states
($\alpha,\beta,\gamma,\delta$) with a variety  of possible two-body scatterings:
\begin{equation}
\chi^\alpha_i + \chi^\beta_j \shortrightarrow \chi^\gamma_i + \chi^\delta_j,  
\end{equation}
where two particles, $i$ and $j$, with initial states $\alpha$ (particle $i$)
and $\beta$ (particle $j$) scatter into two new states $\gamma$ (particle $i$)
and $\delta$ (particle $j$). The particle states have masses
$m^\alpha_i, m^\beta_j$ before the scattering and $m^\gamma_i, m^\delta_j$
after the scattering.  We note that these are simulation particle masses, whereas the actual dark matter particle masses are 
$m_{\chi^\epsilon}$, with $\epsilon=(\alpha,\beta,\gamma,\delta)$,
depending on the state. The velocity-dependent transfer cross section for reaction
${\alpha \beta \shortrightarrow \gamma \delta}$ is given by ${\sigma_{\rm
T}^{\alpha \beta \shortrightarrow \gamma \delta}(v^{\alpha \beta})}$, where $v^{\alpha \beta}$ is the modulus
of the relative velocity between particles in states $\alpha$ and $\beta$. The
scattering rates for the different reaction channels are then given by:
\begin{equation}
R^{\alpha \beta \shortrightarrow \gamma \delta} = \frac{\rho^{\beta}}{m_{\chi^\beta}} \langle\sigma_{\rm T}^{\alpha \beta \shortrightarrow \gamma \delta}(v^{\alpha \beta}) \, v^{\alpha \beta} \rangle,
\end{equation}
where we take the thermal average of the product of the cross section times the relative
velocities between particles. Here $\rho^\beta$ measures the local density of dark matter particles in state $\beta$. We convert this mass density to a number density by dividing it by $m_{\chi^\beta}$, the mass of
the dark matter particle in state $\beta$. During a scattering reaction an energy
$\Delta E^{\alpha \beta \shortrightarrow \gamma \delta}$ is released (exothermic) or
absorbed (endothermic):
\begin{equation}
\Delta E^{\alpha \beta \shortrightarrow \gamma \delta} \left\{
\begin{array}{ll}
=0, & {\rm elastic} ,\\
<0, & {\rm inelastic: endothermic}\\
>0, & {\rm inelastic: exothermic}. 
\end{array}
\right.
\end{equation}
Once particle $i$ in state $\alpha$ and $j$ in state $\beta$ have been selected to scatter into states $\gamma$ and $\delta$, we perform the scattering in the centre of mass frame and assign 
new velocities for the particles after the scattering:
\begin{align}
  {\bf v}_{i}&=\frac{m_i^\alpha + m_j^\beta}{m_i^\gamma + m_j^\delta} {\bf v}_{\rm cm} + \frac{m_j^\delta}{m_i^\gamma + m_j^\delta} \widetilde{v}^{\alpha \beta \shortrightarrow \gamma \delta}_{ij} v_{ij}\,{\bf \hat e},\nonumber\\
  {\bf v}_{j}&=\frac{m_i^\alpha + m_j^\beta}{m_i^\gamma + m_j^\delta} {\bf v}_{\rm cm} - \frac{m_i^\gamma}{m_i^\gamma + m_j^\delta} \widetilde{v}^{\alpha \beta \shortrightarrow \gamma \delta}_{ij} v_{ij}\,{\bf \hat e},
\label{eq:vel}
\end{align}
where ${\bf v}_{\rm cm}$ is the centre of mass velocity of the two particles, $v_{ij}$ their relative velocity, ${\bf \hat e}$ is a random vector
on the unit sphere, and $\widetilde{v}^{\alpha \beta \shortrightarrow \gamma
\delta}_{ij}$ is a dimensionless velocity scale factor that depends on the energy
splitting related to the reaction the two particles are undergoing:
\begin{equation}
\widetilde{v}^{\alpha \beta \shortrightarrow \gamma \delta}_{ij} = \sqrt{\frac{\mu^{\alpha\beta}_{ij}}{\mu^{\gamma\delta}_{ij}}\left(1 + \frac{{2 \Delta E}^{\alpha \beta \shortrightarrow \gamma \delta}}{\mu^{\alpha\beta}_{ij} v_{ij}^2}\right)},
\end{equation}
where ${\mu^{\alpha\beta}_{ij}=m^\alpha_i m^\beta_j/(m^\alpha_i + m^\beta_j)}$ 
and ${\mu^{\gamma\delta}_{ij}=m^\gamma_i m^\delta_j/(m^\gamma_i + m^\delta_j)}$
are the reduced masses of the two particles before and after the scattering event; i.e., we take into account the change in particle mass during the scattering. We note however, that this change is usually tiny in the cases of interest where the mass splitting is small. For example, for our specific nearly-degenerate model, it is negligible ($10^{-6}$) such that the two reduced masses are essentially given by $1/2\,m_0$, assuming a constant dark matter simulation particle mass $m_0$.  Based on the
three cases for $\Delta E^{\alpha \beta \shortrightarrow \gamma \delta}$ we
find accordingly for the dimensionless velocity scale factor:
\begin{equation}
0 \leq \widetilde{v}^{\alpha \beta \shortrightarrow \gamma \delta}_{ij} \left\{
\begin{array}{ll}
=1, & {\rm elastic} ,\\
<1, & {\rm inelastic: endothermic}\\
>1, & {\rm inelastic: exothermic}. 
\end{array}
\right.
\end{equation}
In the inelastic case this energy is either given or taken from the two
scattering particles in equal parts.  The endothermic regime is limited by the
fully inelastic collision ($\widetilde{v}^{\alpha \beta \shortrightarrow
\gamma \delta}_{ij}=0$) after which both particles move with the centre of mass
velocity. The exothermic case is not limited and can in principle inject an arbitrary
amount of energy into the system depending on the energy level splitting.

To decide whether a certain scattering reaction occurs between two simulation particles, we have to evaluate the corresponding scattering probabilities. The pairwise scattering probability for a given reaction ${\alpha \beta \shortrightarrow
\gamma \delta}$, and the total probability for scattering of a given particle with any of its neighbours are given respectively by:
\begin{align}
P^{\alpha \beta \shortrightarrow \gamma \delta}_{ij} &= \delta^{\alpha \beta}_{ij} \,   m_j^\beta \, W_{ij} \, \frac{\sigma_{\rm T}^{\alpha \beta \shortrightarrow \gamma \delta}(v_{ij})}{m_{\chi^\beta}} \, \frac{v_{ij}}{2} \,  {\Delta t_i },\\
P^{\alpha \beta \shortrightarrow \gamma \delta}_i &= \sum_{j=0}^{N_{\rm ngb}} P^{\alpha \beta \shortrightarrow \gamma \delta}_{ij}, 
\end{align}
where $N_{\rm ngb}$ is the number of neighbor particles, $\Delta t_i$ the individual time-step of particle $i$, and ${W_{ij}=w(r_{ij}/h_{i}, h_{i})}$ is the cubic spline Kernel function with a 3D normalisation:
\begin{equation}
w(q,h) =\frac{8}{\pi\,h^3} \left\{
\begin{array}{ll}
1-6 q^2 + 6 q^3,     & 0\le  q \le\frac{1}{2} ,\\
2\left(1-q\right)^3, & \frac{1}{2}< q \le 1 ,\\
0 ,                  & q>1 .
\end{array}
\right.
\end{equation}
Here $h_i$ is the smoothing length including the $N_{\rm ngb}$ neighbor particles $j$ around particle $i$ with spatial distance $r_{ij}$. We note that  $N_{\rm ngb}$ does not distinguish the particle states; i.e., it is
possible that particles of a certain state are not enclosed in the smoothing
length. In this case scattering between the particle in question with this state is not 
possible.  The factor $1/2$ in the calculation of $P^{\alpha \beta
\shortrightarrow \gamma \delta}_{ij}$ accounts for the fact that a scatter
event always involves two particles, and we therefore need to divide by two to
reproduce the correct scattering rate. A given neighbour only contributes to
the sum if the initial states of the reaction match the actual particle states. This is
guaranteed by the Kronecker delta function $\delta^{\alpha \beta}_{ij}$, which is equal to $1$ if particle $i$ is in
initial state $\alpha$ and particle $j$ is in initial
state $\beta$. Otherwise the function evaluates to
zero.

\begin{figure*}
\centering
\includegraphics[width=0.496\textwidth]{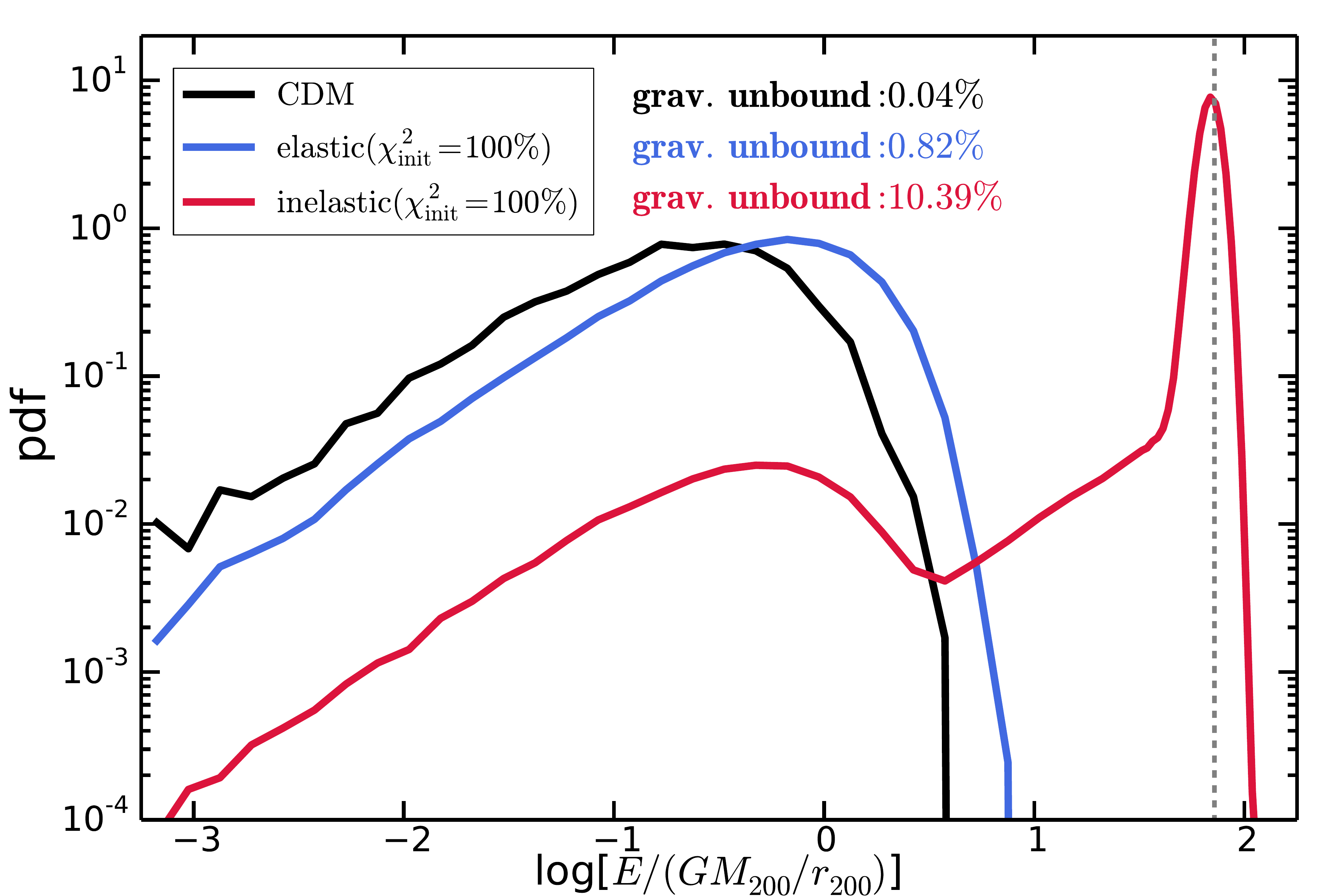}
\includegraphics[width=0.496\textwidth]{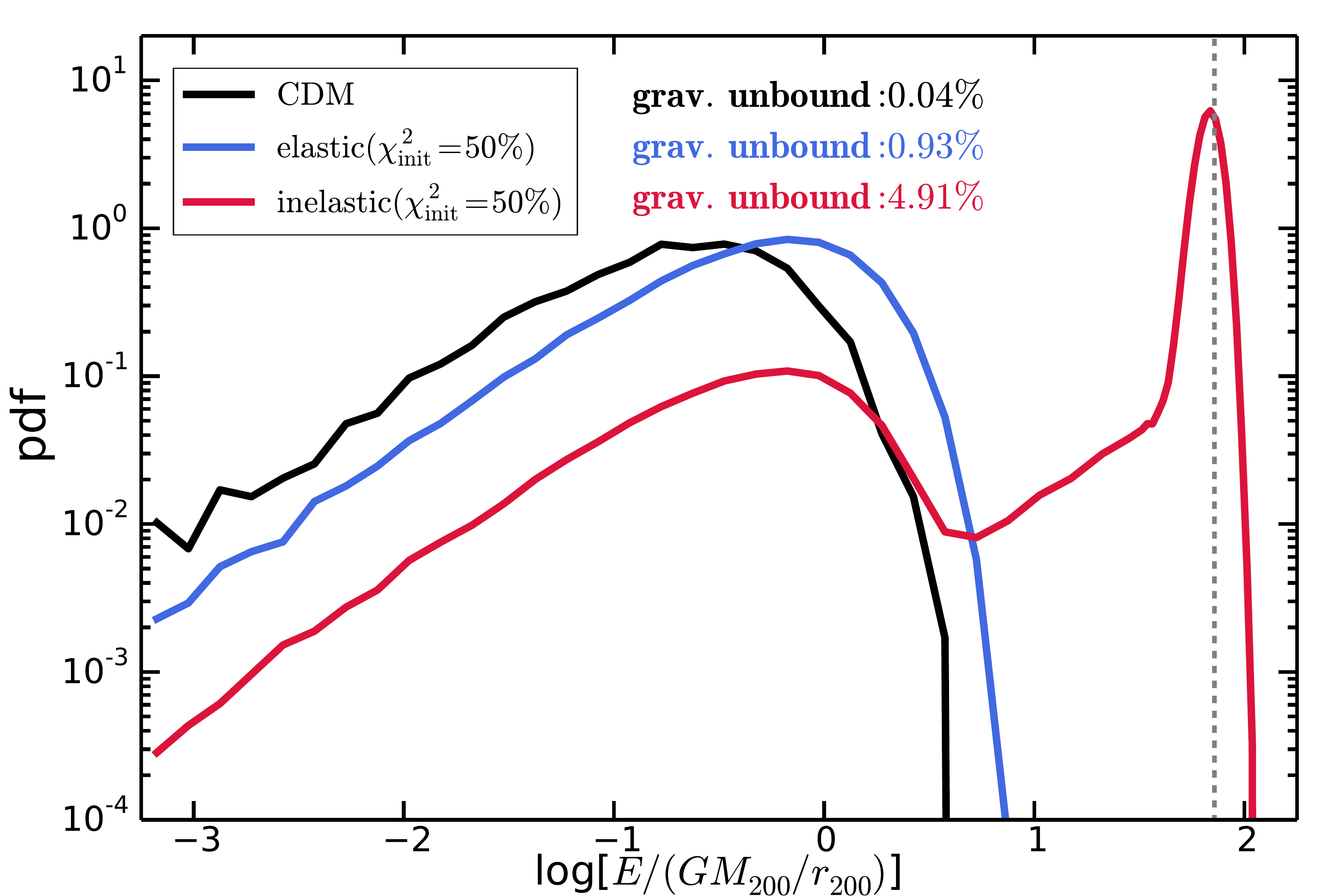}\\
\caption{{\bf Dimensionless energy distribution of gravitationally unbound particles after $\mathbf{10\,{\rm \bf Gyr}}$ for the inelastic models in Fig.~\ref{fig:level_evolution}.}
In the {\it left panel}, $100\%$ of the particles are initially in the excited state,
while on the {\it right panel}, $50\%$ of the particles are initially excited and the rest is in the ground
state. The legends on the panels also show the total fraction of gravitationally unbound particles in the
different models corresponding to the black (CDM), blue (elastic) and red (inelastic) lines.
The inelastic SIDM model leads to a significant removal of particles, predominantly from the halo centre,
and has a peak at an energy around ${\sim 74\,G M_{\rm 200}/r_{\rm 200}}$ due to the exothermic reactions (gray, dashed vertical line; see text for an analytical estimate). The elastic SIDM model has a distribution
similar to the CDM distribution, but slightly shifted towards larger energies. The fraction of gravitationally unbound particles
is about twice as large for the $\ICfull$ initial configuration compared to the
$\IChalf$ case.}
\label{fig:energy_hist}
\end{figure*}

To decide whether a particle is going to scatter and with which reaction, we
first arbitrarily order the reactions according to ${0 \leq \zeta = (\alpha
\beta \shortrightarrow \gamma \delta) \leq \zeta_{\max}}$, where $\zeta_{\max}$
denotes the total number of reactions.  For each particle we then draw a random
number ${x_i^P\in U(0,1)}$. A scattering of particle $i$ occurs if there is a
reaction ${\zeta_i \leq \zeta_{\rm max}}$ with 
\begin{equation}
\sum_{\zeta=0}^{\zeta_i-1} P^\zeta_i < x_i^P  < \sum_{\zeta=0}^{\zeta_i} P^\zeta_i.
\end{equation} 
If such a $\zeta_i$ exists, particle $i$ will scatter with reaction ${\zeta_i =
(\alpha_i \beta_i \shortrightarrow \gamma_i \delta_i)}$. This then determines
the energetics of the scattering process (inelastic, exothermic or
endothermic), the state of the scattering partner ($\beta_i$), and the final
states of the reaction ($\gamma_i, \delta_i$). Once the scattering reaction is decided, a scattering partner $j_i$ for particle
$i$ needs to be found.  The exact partner depends on the reaction since it must
be a particle which is currently in state $\beta_i$ and will then scatter into
state $\delta_i$. The selection is done by finding the partner $j_i$ that satisfies:  
\begin{equation}
P^{\zeta < \zeta_i}_i + \sum_{j=0}^{j_i-1} P^{\zeta_i}_{ij} < x_i^P < P^{\zeta < \zeta_i}_i + \sum_{j=0}^{j_i} P^{\zeta_i}_{ij},
\end{equation}
where $P^{\zeta < \zeta_i}$ is the sum of all probabilities for reactions smaller than $\zeta_i$, i.e. $P^{\zeta < \zeta_i} = \sum_{\zeta=0}^{\zeta_i-1} P^{\zeta}_i$.

Once the scattering partner $j_i$ is found, the scattering can be performed by
assigning new velocities to particle $i$ and $j_i$ based on the dimensionless
velocity scale factor $\widetilde{v}^{\zeta_i}_{i j_i}= \widetilde{v}^{\alpha_i
\beta_i \shortrightarrow \gamma_i \delta_i}_{i j_i}$. In a last step we assign
the new states to the scattering particles, i.e., the state of particle $i$ is
changed to $\gamma_i$ and the state of particle $j_i$ is changed to $\delta_i$.
We also change the masses of the particles to reflect their state change, although we note that in our simulation this only results in a minimal mass
change due to the small mass splitting for the nearly degenerate system presented above. 

To avoid multiple scattering per particle in a single time-step, we impose a limit to the time-step
of each particle $i$:
\begin{equation}
\Delta t_i < \kappa \min_{\alpha, \beta, \gamma, \delta}(\rho_i^\alpha \sigma_{\rm T}^{\alpha \beta \shortrightarrow \gamma \delta}(\sigma_i^\alpha)/m_{\chi^\alpha} \sigma_i^\alpha)^{-1}, 
\end{equation}
where $\rho_i^\alpha$ is the density of particles in state $\alpha$ at the
location of particle $i$, and $\sigma_i^\alpha$ is the corresponding velocity
dispersion. $\kappa$ is a dimensionless parameter that can be adjusted to
control the size of the minimal time-step. For our simulations we find that $\kappa=0.0025$
is sufficient to avoid multiple scattering and usually sets the time-step to be smaller than the
local dynamical time scale in the simulations.

We test our inelastic SIDM implementation by setting up 
an isolated~\cite{Hernquist1990} halo in equilibrium with ${M_{\rm 200}=10^{10}\msun}$ and a
concentration\footnote{Defined as the ratio of $r_{\rm 200}$, the radius where the mean density of the
halo is equal to $200$ times the critical density, and $r_{-2}$, the radius where the logarithmic slope of the 
density profile is equal to $-2$.} of $c=8$ for the benchmark dark matter model presented above. The
halo is sampled with $10^7$ particles and gravitational forces are softened
with a Plummer-equivalent softening length of ${\epsilon=10\pc}$. 
We simulate
this halo in three different dark matter models: CDM, elastic SIDM and inelastic SIDM.
The CDM case just follows the evolution of collisionless CDM, whereas the other
two cases employ the two-state model with the cross sections discussed above (see bottom panel of Fig.~\ref{fig:schematic}) using
the numerical implementation described in the previous section. The elastic
simulation artificially suppresses the energy release during (de-)excitations,
but is otherwise identical to the inelastic SIDM model.  This simulation can
then be compared to the inelastic case, where velocity kicks
play a relevant role. 
To be more specific, the elastic case is realised by simply setting
${\widetilde{v}^{\alpha \beta \shortrightarrow \gamma \delta}_{ij}=1}$ during
any scattering process, i.e., ${\Delta E^{\alpha \beta \shortrightarrow \gamma
\delta}=0}$.
As mentioned above, we explore two different initial configurations for the
non-CDM models. In the first configuration all particles are initially in the
excited state ($\ICfull$), while in the second
only half of the particles are initially in the excited state
($\IChalf$).

\begin{table*}
\begin{center}
\begin{tabular}{lccccccc}
\hline
Model                   & $M_{\rm 200}$          & $r_{\rm 200}$    & $V_{\rm max}$                   & $R_{\rm max}$  & $N_{\rm sub}$  & ground state population                 & injected energy\\
                        & $[10^{12}\,{\rm M_\odot}]$  & $[{\rm kpc}]$         & $[{\rm km}\,{\rm s}^{-1}]$      & $[{\rm kpc}]$  &                & $[\%]$ $(r < 300\kpc)$                  & $[$SNIIs $(10^{51}\,{\rm erg})]$   \\
\hline
\hline
CDM                                                          & $1.609$ & $243.85$ & $174.87$ & $66.14$ & $16,264$ & --    & --     \\
\hline
elastic SIDM ($\ICfull$)                                     & $1.600$ & $243.39$ & $178.12$ & $62.22$ & $14,486$ & $7.51$  & --     \\[0.1cm]
elastic SIDM ($\IChalf$)                                     & $1.600$ & $243.38$ & $177.51$ & $63.39$ & $14,675$ & $52.23$ & --     \\
\hline
inelastic SIDM ($\ICfull$)                                   & $1.478$ & $237.05$ & $164.14$ & $77.21$ & $13,171$ & $0.65$  & $755 \times 10^6$ \\[0.1cm]
inelastic SIDM ($\IChalf$)                                   & $1.569$ & $241.81$ & $172.44$ & $64.76$ & $14,409$ & $51.12$ & $268 \times 10^6$ \\
\hline
\end{tabular}
\end{center}
\caption{{\bf Basic properties of the Milky Way-sized halo simulated in the
different dark matter models.} We list the virial mass ($M_{\rm 200}$), virial
radius ($r_{\rm 200}$), maximum circular velocity ($V_{\rm max}$), radius
where the maximum circular velocity is reached ($R_{\rm max}$), the number of
resolved subhaloes within $300\kpc$ ($N_{\rm sub}$), the ground state
population of all dark matter particles within $300\kpc$, and the injected energy (in
units of ${10^{51}\erg}$, i.e., the canonical energy of a SNII) measured based on the
population split between the two dark matter states of all high resolution particles at $z=0$.} \label{table:halo}
\end{table*}

In Fig.~\ref{fig:level_evolution} we present the time evolution of
the population split in the halo for the two initial conditions for the inelastic SIDM case,
in the left and right panels, respectively. The right axis in each panel 
indicates the total injected energy into the system (solid white
lines) in units of the canonical SNII energy ($10^{51}\,{\rm erg}$). This
measures the total energy released due to de-excitation that is 
transformed into kinetic energy by the cumulative effect of velocity 
kicks in each scattering event.
As expected, in both cases the ground state population increases over time while the excited state gets
de-populated. As mentioned above, this is a consequence of the employed cross
sections, where up-scattering is strongly suppressed  (see bottom panel of Fig.~\ref{fig:schematic}), and does not occur given
the typical relative velocities of particles in a $10^{10}\msun$ halo. After a few
${\rm Gyr}$, the exothermic reaction has already injected the equivalent of more than one
million SNII for the $\ICfull$ configuration.
After $10\Gyr$, the cumulative energy injection of down-scatterings has reached $\sim 2\times 10^{57} \erg$ of
energy into the system for the $\ICfull$ 
configuration, and $\sim 7 \times 10^{56} \erg$ for the $\IChalf$ configuration. 
The ground state population increases from
$0\%$ to $10\%$ for the $\ICfull$ initial
configuration over the simulation time span of $10\Gyr$. Similarly also the
ground state population for the $\IChalf$ initial
configuration increases as a function of time, although the total increase is
lower in this case.  As a reference, we note that the minimum energy required to transform
a cusp into a 1~kpc core for a Navarro-Frenk-White (NFW) halo with a mass of $10^{10}\msun$ calculated
from the virial theorem assuming initial and final equilibrium configuration is $\sim10^{55}\erg$ \citep{Penarrubia2012}.
Depending on the stellar mass content, efficiency of energy injection, and star formation history of a given galaxy living in a halo of this size, this energetic requirement might or might not be satisfied by the SNe-driven core formation scenario \citep[e.g.][]{Penarrubia2012,Amorisco2014,Maxwell2015}. In the inelastic SIDM model explored here the cumulative energy 
available easily exceeds this minimum energy requirement.

The mass splitting, which corresponds to a velocity kick of $424\kms$ is
sufficient to unbind particles, i.e., to efficiently remove dark matter particles
from the $10^{10}\msun$ halo. This is demonstrated in Fig.~\ref{fig:energy_hist}, where we show
the dimensionless energy distribution of all gravitationally unbound particles after $10\Gyr$. The CDM case (black lines)
has a very small fraction of gravitationally unbound particles $\mathcal{O}(0.01\%)$, caused by numerical
noise over the equilibrium configuration. This population serves as a comparison baseline with
the other models. On the other hand, the inelastic model (red lines) has a much larger fraction of gravitationally unbound particles
($\sim5\%-10\%$), 
with a significant number of them
populating a nearly log-normally distributed peak at high energies. 
These are the particles that were predominantly expelled from the halo centre
during the velocity kicks imparted in down-scatterings.
Notice that this population is absent in the elastic case (blue lines),
which shows a very similar distribution as the CDM case, but shifted towards
higher energies and with a higher fraction of gravitationally unbound particles, of $\mathcal{O}(1\%)$. This distinct population of gravitationally unbound particles in the inelastic case
is key to understand the further reduction of the core density that happens in this case
compared to pure elastic SIDM. 

\begin{figure}
\centering
\includegraphics[width=0.495\textwidth]{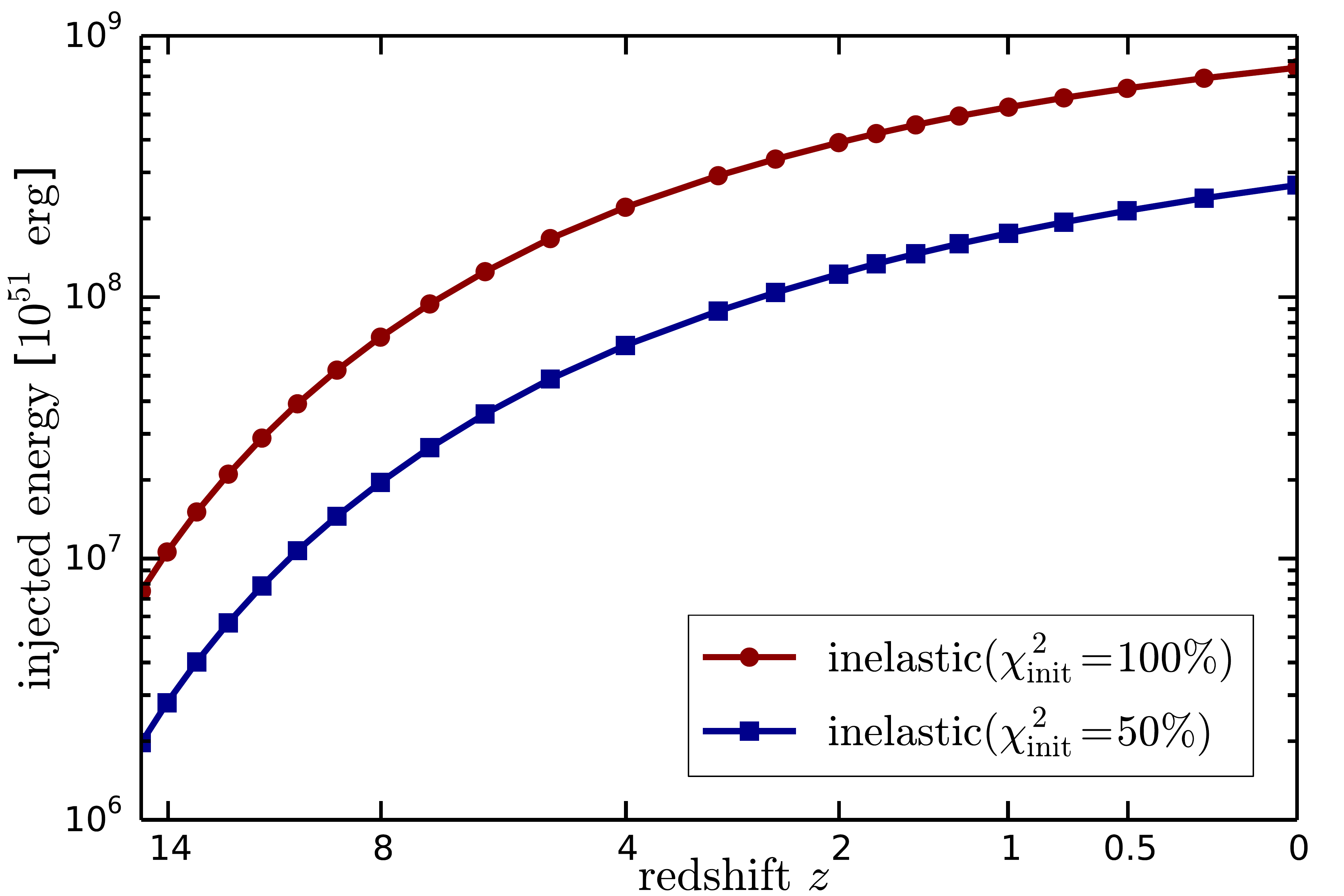}
\caption{{\bf Injected energy into the Milky Way-like halo due to level
de-excitation.} For both initial configurations ($\ICfull$ and
$\IChalf$) the total injected energy into the halo due to down-scattering of the excited
states is equivalent to hundreds of millions of SNII ($10^{51}\erg$). This energy
leads to an increased core formation compared to elastic SIDM models.
}
\label{fig:injected_energy}
\end{figure}
\begin{figure*}
\centering
\includegraphics[width=0.495\textwidth]{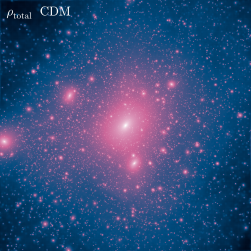}
\includegraphics[width=0.495\textwidth]{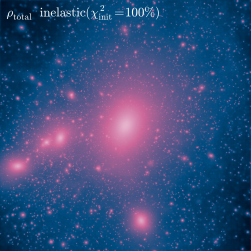}
\caption{{\bf Projected dark matter density for the Milky Way-sized halo in CDM and for inelastic SIDM.} The {\it left panel} shows the CDM case,
whereas in the {\it right panel} we present the inelastic SIDM case  where
$100\%$ of the particles are initially in the excited state ($\ICfull$). The
spatial extent of the maps is $500\kpc$ across with a depth of $300\kpc$, with
a colour scale related to the projected value of $\rho^2$. It is apparent that
the inelastic SIDM model leads to reduced core densities and an overall
reduction in the abundance of subhaloes due to down-scattering. The virial mass
for the halo in the inelastic case is reduced by about $10\%$ compared
to the CDM case due to the removal of dark matter particles in the ground state
following de-excitation.  The efficient removal of ground state particles leads
to a very small fraction, $\lesssim 1\%$, of such particles within $300\kpc$.}
\label{fig:dmtotal_cdm_map}
\end{figure*}

We can roughly estimate the energy shifts in the elastic and inelastic SIDM cases. For
the elastic case we assume that the shift seen between the blue and black lines is related to the energy gained 
by the particles that were barely bound in the inner halo, and that become gravitationally unbound
after elastic scattering with particles with more kinetic energy. Since scattering is more
frequent deep within the potential, this energy gain is approximately bounded by the
the r.m.s. velocity at the maximum of the velocity dispersion profile, which occurs at
$r_{\sigma, \rm max} \sim a/5$, where $a$ is the scale radius of the Hernquist profile.
In~\cite{Vogelsberger2012} we found that the r.m.s. velocity at the core is ${v_{\rm rms}^2 (r_{\sigma, \rm max}) \sim 3 \sigma^2 (r_{\sigma, \rm max}) \sim 0.96\,GM_{\rm 200}/a}$ (see Fig. 2 of~\citealt{Vogelsberger2012}). This
means that those particles that were barely bound become gravitationally unbound gaining an additional
kinetic energy $0.48\,GM_{\rm 200}/a\sim3.42\, GM_{\rm 200}/r_{\rm 200}$, which 
is roughly 0.5 dex to the right, relative to the CDM distribution where this effect is absent.
We note that our analytic estimate is an oversimplification given the radially
dependent re-distribution of energy.
For the inelastic case the elastic collisions still cause a similar shift as
in the elastic case since the reaction (${\chi^1 + \chi^2 \rightarrow \chi^1
+ \chi^2}$) is very frequent (it has the highest cross section). In addition, during
down-scatterings, the resulting ground state particles receive a velocity
kick $\sqrt{2\delta/m_{\chi^1}}$, which results in a gain in
kinetic energy of $\sim 74\,GM_{\rm 200}/r_{\rm 200}$, which is shown in Fig.~\ref{fig:energy_hist} as a vertical dashed line. This 
estimate describes the numerical result well, demonstrating that the numerical implementation behaves correctly.

\section{Impact of inelastic SIDM on a Milky Way-like halo}

\begin{figure*}
\centering
\includegraphics[width=1\textwidth]{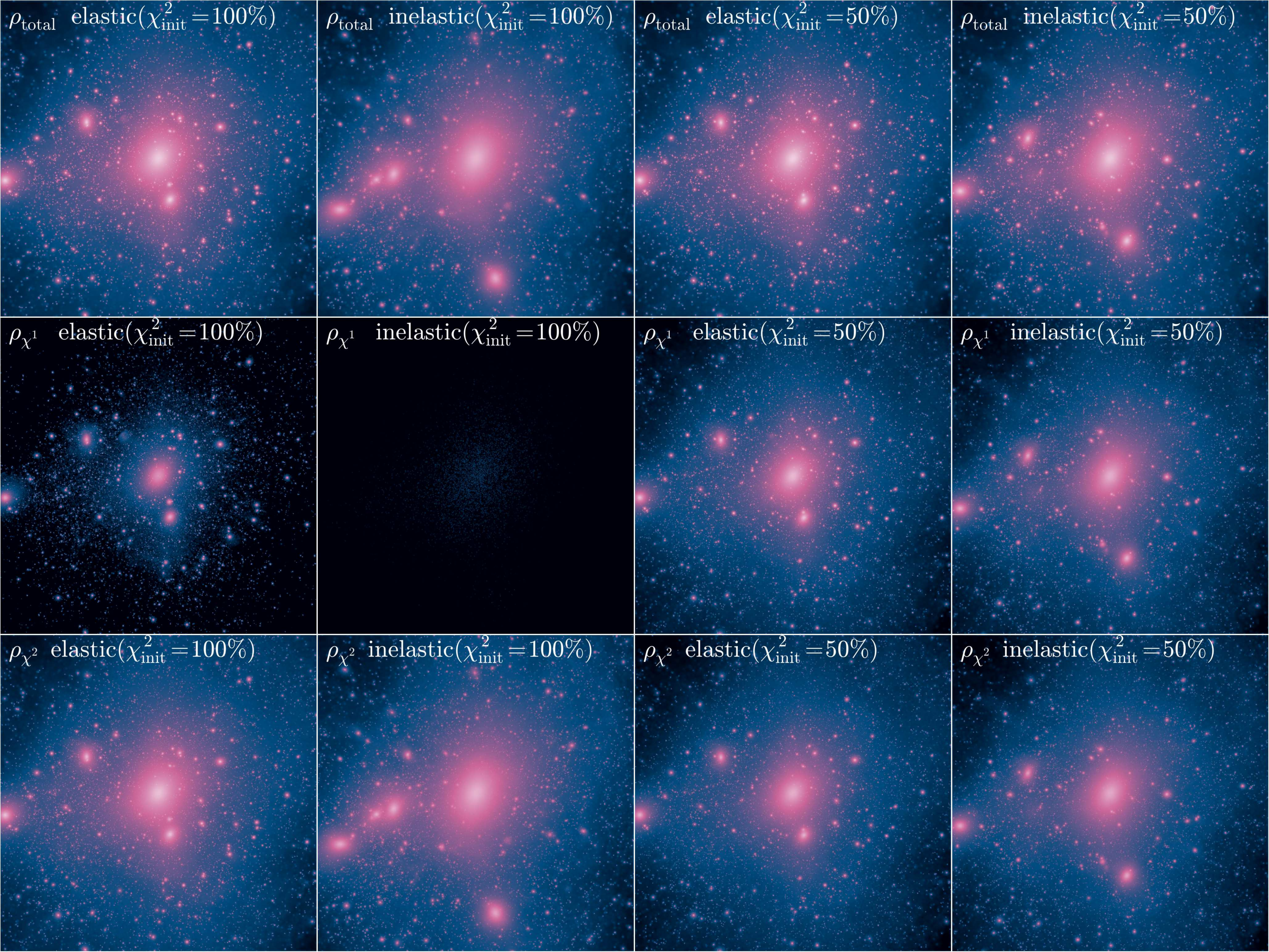}
\caption{{\bf Projected dark matter density for the total dark matter density, the lower level, and the
upper level densities for the SIDM models.} The total extent of the maps is
$500\kpc$ with a depth of $300\kpc$, and we project $\rho^2$. {\it Top row
panels:} Total dark matter density fields for the halo in different SIDM models as
indicated. The largest difference compared to the CDM case occurs for the
inelastic model with the $\ICfull$ initial
configuration.  The reduction of core densities and subhalo abundance is smaller for the $\IChalf$ initial configuration compared to the $\ICfull$ case. {\it Middle row panels:} Dark matter densities of the ground state
($\chi^1$) for the different models. For the inelastic
case nearly all excited particles have been removed from
the halo centre due to the energy injection during de-excitation of the 
state. This is not the case for the elastic model, where the ground
state particles stay close to the halo centre and are only redistributed due to
scattering events. {\it Bottom row panels:} Dark matter densities of the excited state ($\chi^2$) for the different models.}
\label{fig:dm_maps}
\end{figure*}

Next we explore the impact of inelastic SIDM on a galactic halo in a
cosmological context. We resimulated the galactic halo
($M_{\rm 200} = {1.6\times10^{12}\msun}$) presented in~\cite{ETHOS2} within inelastic SIDM for
the benchmark model presented above with two different initialisations,
$\ICfull$    and $\IChalf$.  The simulations employ the following cosmological
parameters:  ${\Omega_m=0.302}$, ${\Omega_{\Lambda}=0.698}$, ${\Omega_b=0.046}$,
${h=0.69}$, ${\sigma_8=0.839}$ and ${n_{s}=0.967}$, which are consistent with recent
Planck data~\citep[][]{Planck2014, Spergel2015}. The initial conditions are generated at $z=127$. The gravitational softening
length is fixed in comoving coordinates until $z=9$, and is then fixed in
physical units until $z=0$.  The dark matter particle mass resolution is  $ 2.756\times
10^4\msun$ with a Plummer-equivalent softening length of $72.4\pc$ at $z=0$.
A convergence study is presented in Appendix~A.

The main properties of this halo are presented in Table~\ref{table:halo} for
the four different SIDM models, elastic and inelastic with the two different
initial state populations explored in this paper.  We can see that the virial
mass, $M_{\rm 200}$, is reduced by nearly $\sim 10\%$ for the inelastic model
with initially all particles in the excited state ($\ICfull$). This mass loss
is a consequence of unbinding ground state particles after down-scattering. We
also find a rather significant reduction in $V_{\rm max}$ of about $\sim
8.5\%$. The total number of resolved subhaloes within $300\kpc$ is also reduced
by $\sim 23\%$ for the inelastic model with $\ICfull$. Here we count all subhaloes that are found by our structure finder~\citep[][]{Springel2001} with more than $20$
particles corresponding to a lower mass limit of $5.512 \times 10^5\msun$. We
note that there is a significant reduction in the abundance of subhaloes in
both the elastic and inelastic SIDM model. The fact that this also happens for the
elastic model is caused  by the rather large elastic cross sections
that we employ in our benchmark model (see bottom panel of Fig.~\ref{fig:schematic}). In
particular, the cross section for the (${\chi^1 + \chi^2 \rightarrow \chi^1 +
\chi^2}$) reaction reaches values larger than $10\cpm$ for low relative
velocities, and it is around $5\cpm$ at relative velocities $\sim200\kms$,
which are the relevant ones for subhalo evaporation. For instance, in
\cite{Vogelsberger2012} it was demonstrated that an elastic cross section of
$\sim 10\cpm$ on galactic scales leads to a significant suppression of
substructure. As anticipated, this reduction in the abundance of subhaloes is
clearly enhanced if inelastic scattering is included.  As is clear from
Table~\ref{table:halo}, qualitatively similar trends are observed for the
$\IChalf$ configuration.  The second-to-last column of Table~\ref{table:halo}
shows the present-day population of ground state particles within a radius of
$300\kpc$ from the galactic centre. We find that for the inelastic model the
ground state population increases to $\sim 0.7\%$ for the $\ICfull$ initial
configuration and to $\sim 51.1\%$ for the $\IChalf$ initial configuration. On
the other hand, for the elastic case we find larger fractions of particles in
the ground state at $z=0$ than in the inelastic case,  $\sim 7.5\%$ and $\sim
52.2\%$ for the $\ICfull$ and $\IChalf$ configurations, respectively. This is
because the elastic configuration has been set on purpose to have the same
reactions as the inelastic benchmark case, but without the energy release (see
Section \ref{sec:numerics}),  i.e., there is no velocity kick associated to
down-scattering, which means that the associated ground state particles remain
bound to the halo. For the inelastic case, we provide in the last column of the
table the cumulative injected energy due to de-excitations.  For the fully
excited initial configuration we find that a total of $\sim 8\times10^{8}
\times 10^{51}\erg$ are injected.  This value is a factor of a few lower ($\sim
3\times10^{8} \times 10^{51}\erg$) for the configuration where only $50\%$ of
the particles are initially in the upper state. The redshift dependence of the injected energy
is presented in Fig.~\ref{fig:injected_energy}.
The magnitude of these energies suggest that inelastic SIDM can have a substantial impact on the
galactic halo, both in terms of abundance of substructure, and their density structure.

The projected dark matter density distribution of the simulated halo for a CDM universe
is shown in the left panel of Fig.~\ref{fig:dmtotal_cdm_map}, while the
inelastic SIDM model with $\ICfull$ is shown in the right panel. These plots
clearly show that in this specific inelastic SIDM model, both the central
(sub)halo densities and subhalo abundance are significantly reduced.  This
latter property is quite distinct from typical elastic SIDM models where a
relevant difference in the subhalo abundance relative to CDM is only possible
for rather large elastic cross sections on galactic scales ($\sim 10\cpm$). 

The inelastic SIDM model with $\ICfull$ represents the most extreme scenario of
all our simulations. To show the larger diversity of the non-CDM cases we
explored, we present the corresponding maps of all of them in the four top
panels of Fig.~\ref{fig:dm_maps}.  These four panels show the total dark matter density,
i.e., taking into account the ground and excited states.  Overall, we see a
consistent trend on the abundance of substructure being more suppressed in the
inelastic cases compared to the elastic ones.  The effect is however,
considerably stronger in the case where all particles are initially in the
excited state (the two leftmost panels are visually distinct, while the
rightmost panels are more alike).  In the middle and bottom panels of
Fig.~\ref{fig:dm_maps} we show the density maps but only considering particles
in the ground state, and excited state, respectively.  The inelastic model with
$\ICfull$ has a very low ground state density. De-excited particles escape from the halo
centres in that model due to the strong velocity kicks.  For the associated
elastic case, these particles are not removed and therefore stay
near the halo centres and are visible in the maps.  A visual comparison between
the two leftmost panels of Fig.~\ref{fig:dm_maps} shows very clearly the
striking difference between elastic and inelastic SIDM models.  For the
$\IChalf$ configuration, the differences between the ground and excited state
populations, and between the elastic and inelastic cases are barely visible in
the maps since the increase of ground state particles due to de-excitations is
only at the per cent level.

\begin{figure*}
\centering
\includegraphics[width=0.495\textwidth]{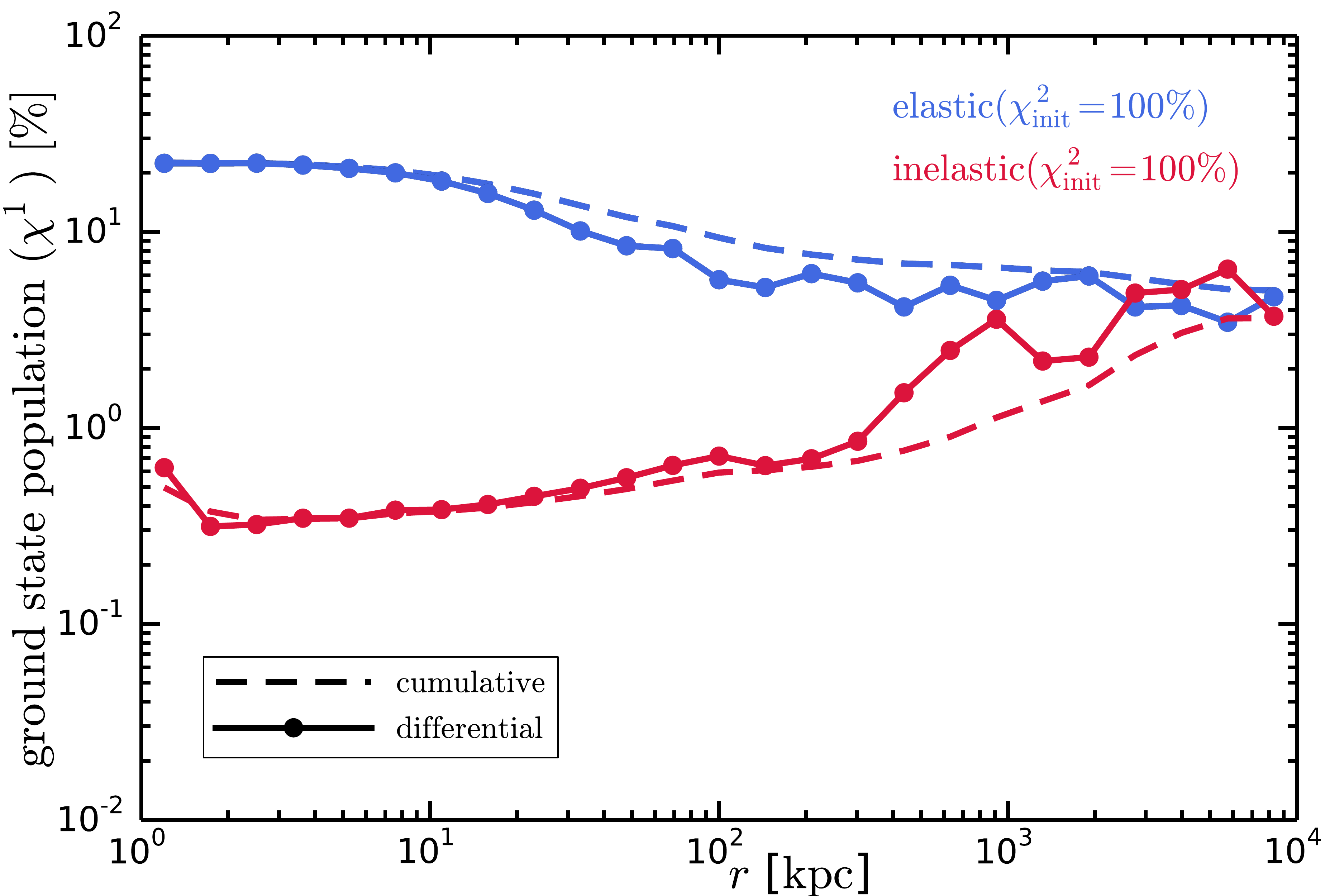}
\includegraphics[width=0.495\textwidth]{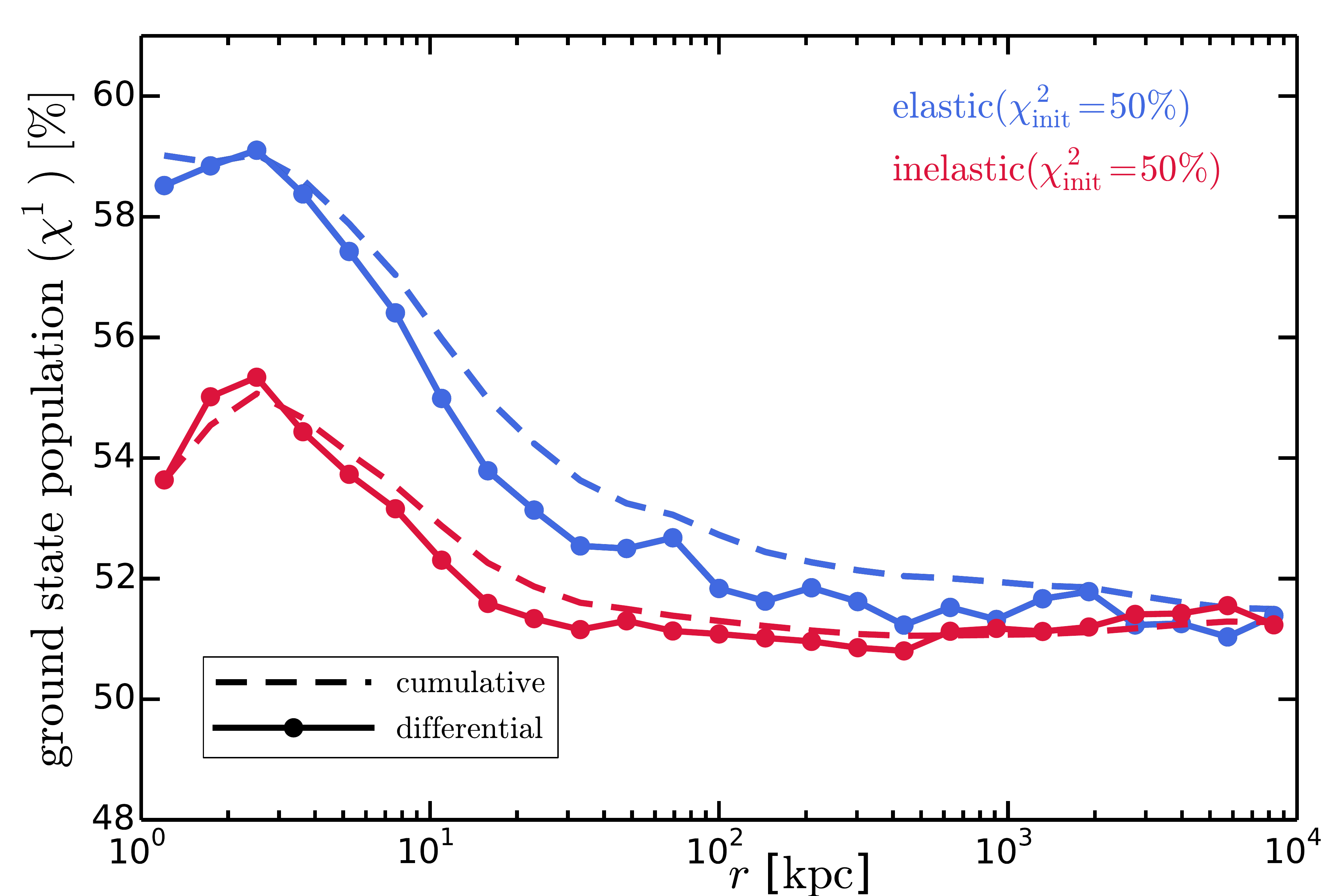}\\
\includegraphics[width=0.495\textwidth]{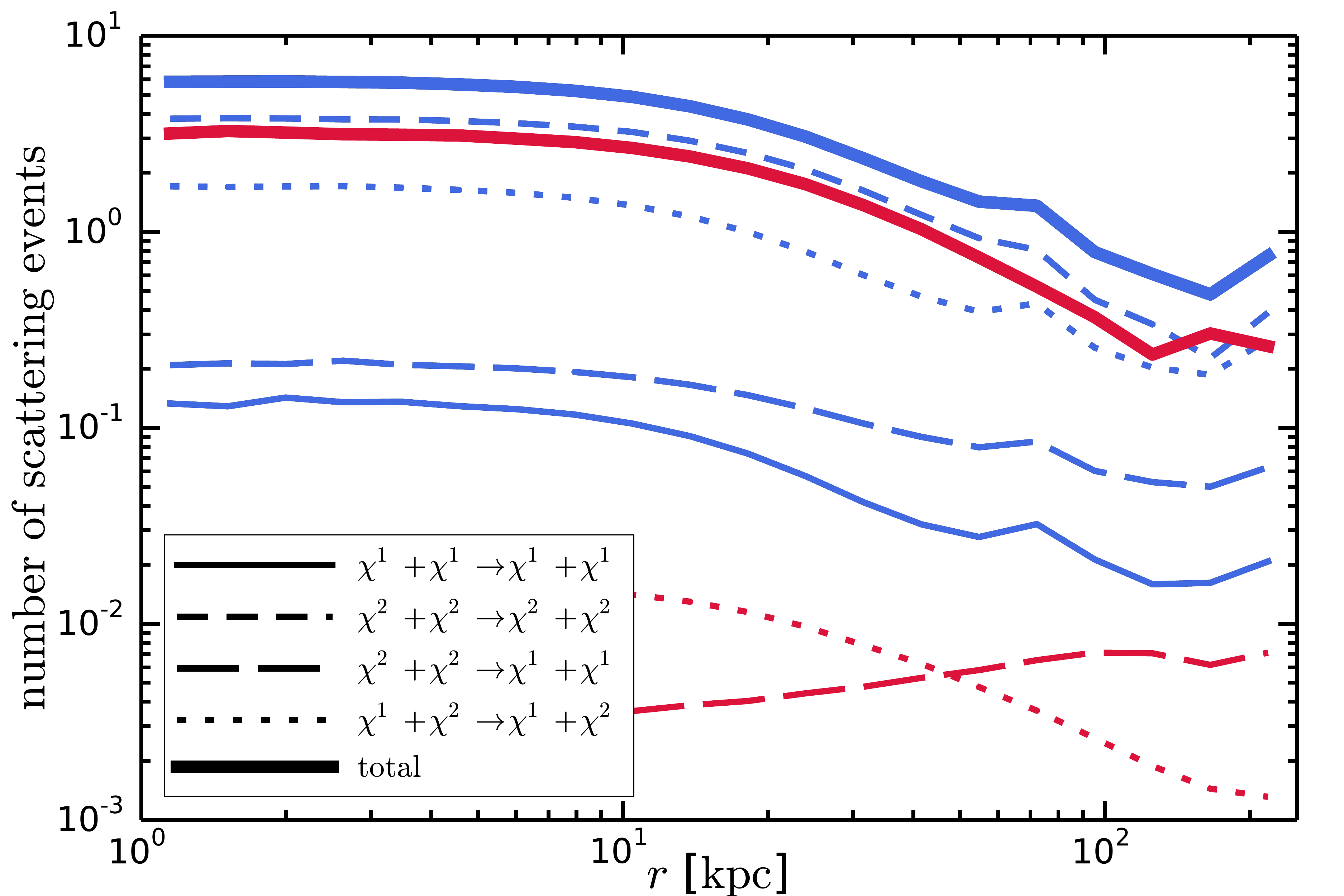}
\includegraphics[width=0.495\textwidth]{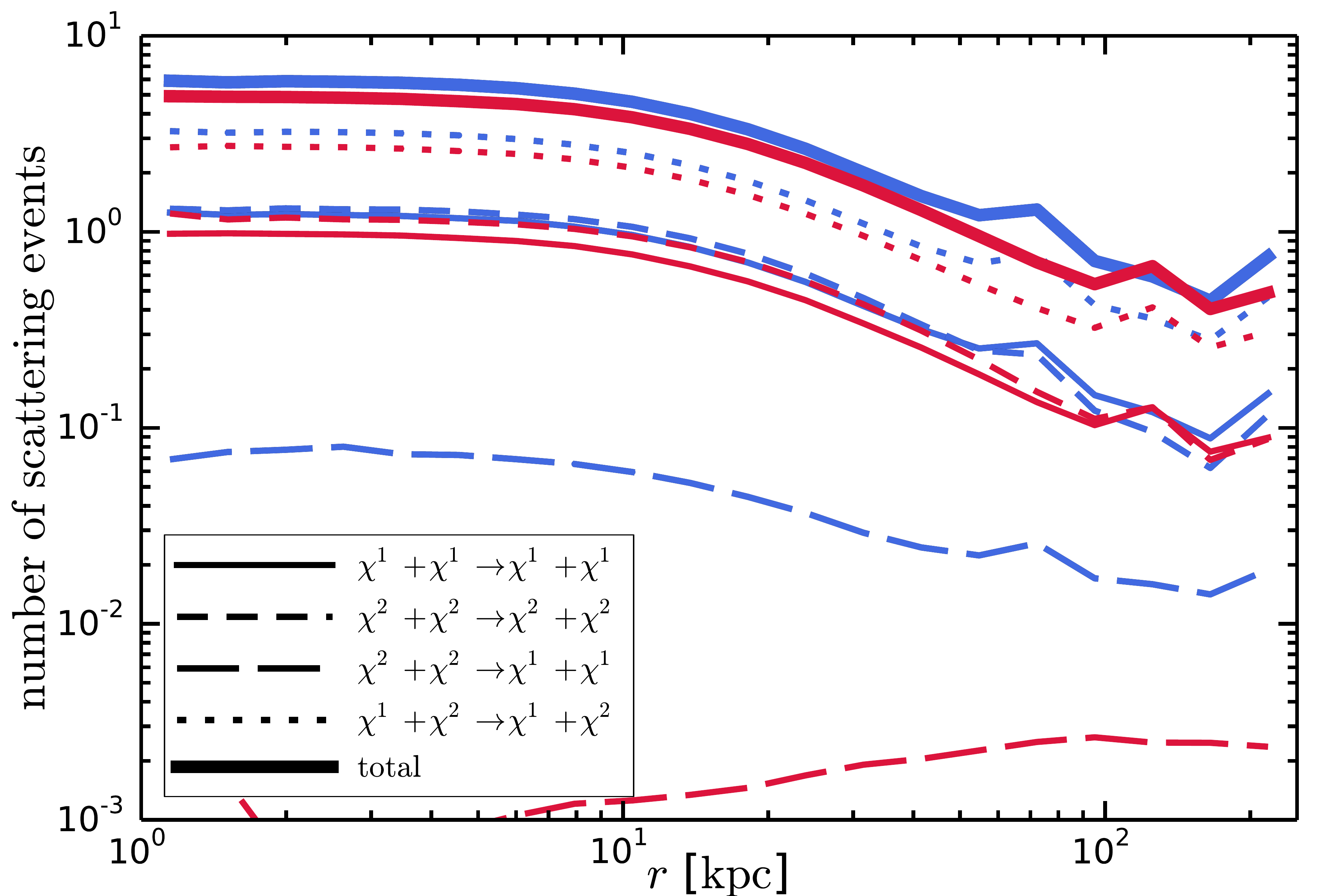}\\
\caption{{\bf Level population and scatter statistics of the Milky Way-sized halo.} {\it Top panels:} Radial
profiles of the ground state population for the $\ICfull$ configuration in the left, and $\IChalf$ in the right
(solid: differential, dashed: cumulative). We show the
elastic and inelastic cases with blue and red lines, respectively. For the inelastic case, down-scattered
particles escape the centre of the halo, which causes a strong suppression of
ground state particles. {\it Bottom panels:} Radial profiles of the mass-weighted average number of scattering
events in each reaction channel according to the legend. The number of scattering events in each channel
depends on the initial configuration and the cross section of each channel. 
For the $\IChalf$ case (right panel), the
distribution of scattering events reflects the ranking of
the cross section (see bottom panel of Fig.~\ref{fig:schematic}). Comparatively, this is different for the $\ICfull$ case (left panel),
where the reactions (${\chi^1 + \chi^2 \rightarrow \chi^1 + \chi^2}$) and
(${\chi^1 + \chi^1 \rightarrow \chi^1 + \chi^1}$)
are suppressed since no ground state scattering partners are available initially.
The thick solid lines show the total number
of scatters summed over all channels. We note that we do not show reactions with less
then $10^{-3}$ average scattering events. Within the relevant radial range, i.e., within $r=R_{\rm max}\sim60$~kpc,
all the cases have $\mathcal{O}$(1) scattering events per particle (including all channels) by $z=0$.
}
\label{fig:frac_profiles}
\end{figure*}

To quantify in detail the distribution of the two-state population in the SIDM
models, we show in the top panels of Fig.~\ref{fig:frac_profiles} the radial
profiles of the ground state population centered in the Milky Way-sized halo.  The
differences between the central distributions of the ground and excited state
populations are striking, particularly for the case where all particles are
initially in the excited state ($\ICfull$, left panel).  As mentioned above this is due to the large energy imparted to the ground state
particles during down-scattering, which is large enough to remove these
particles from the halo.  These removed particles are actually visible in
Fig.~\ref{fig:frac_profiles} far away from the halo centre in the differential
ground state population distribution. They appear as peaks in this distribution
around and beyond $1\Mpc$ from the halo centre. Since the velocity kick in our
benchmark model is $424\kms$, the particle would travel $\sim1\Mpc$ already
within $3\Gyr$.  For the elastic case on the  the other hand, the ground state
population rises towards the centre since the down-scattering rate is higher
towards the denser and colder halo centre (see bottom panel of Fig.~\ref{fig:schematic}), unimpeded
due to the lack of energy release.  Here we find that the ground state
population reaches more than $20\%$ towards the centre of the halo. Even at a
distance of $\sim 100\kpc$ we still find a cumulative fraction of $\sim10\%$ of
all particles in the ground state for the elastic mode, whereas for the
inelastic case, this fraction is much smaller and less than a per cent even out
to $\sim 1\Mpc$. The right panel of Fig.~\ref{fig:frac_profiles} shows the
profiles for the case where $50\%$ of all particles are initially in the
excited state ($\IChalf$). We find here a similar trend when comparing the
elastic and inelastic cases, i.e., the ground state is more densely populated
for the elastic case in the inner halo.  For the elastic case about $55\%$ of
all particles in the centre are in the ground state. Lastly, we note that the
cumulative fraction of ground state particles within $\sim 10\Mpc$ agrees
between the elastic and inelastic cases for both initial state configurations.
This population is at the per cent level above the initial ground state
population. This radial distance is large enough to include even those
particles that were ejected due to de-excitation in the inelastic SIDM model,
and thus, the ground state level populations converge to similar values at
large radii. Any remaining deviations are due to differences in the detailed
scatter reactions that occurred during the assembly history of the halo.

\begin{figure*}
\centering
\includegraphics[width=0.495\textwidth]{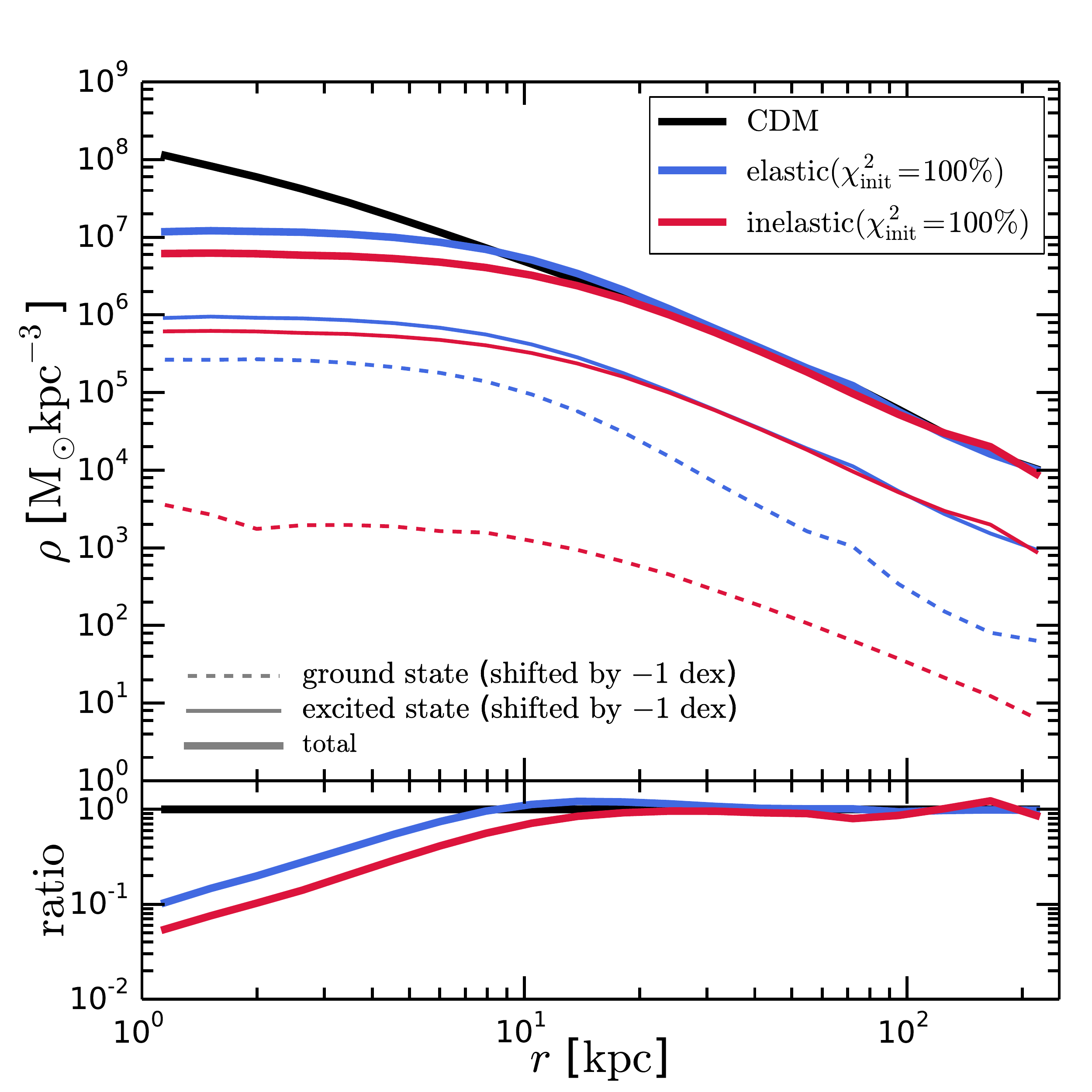}
\hspace{0.11cm}\includegraphics[width=0.495\textwidth]{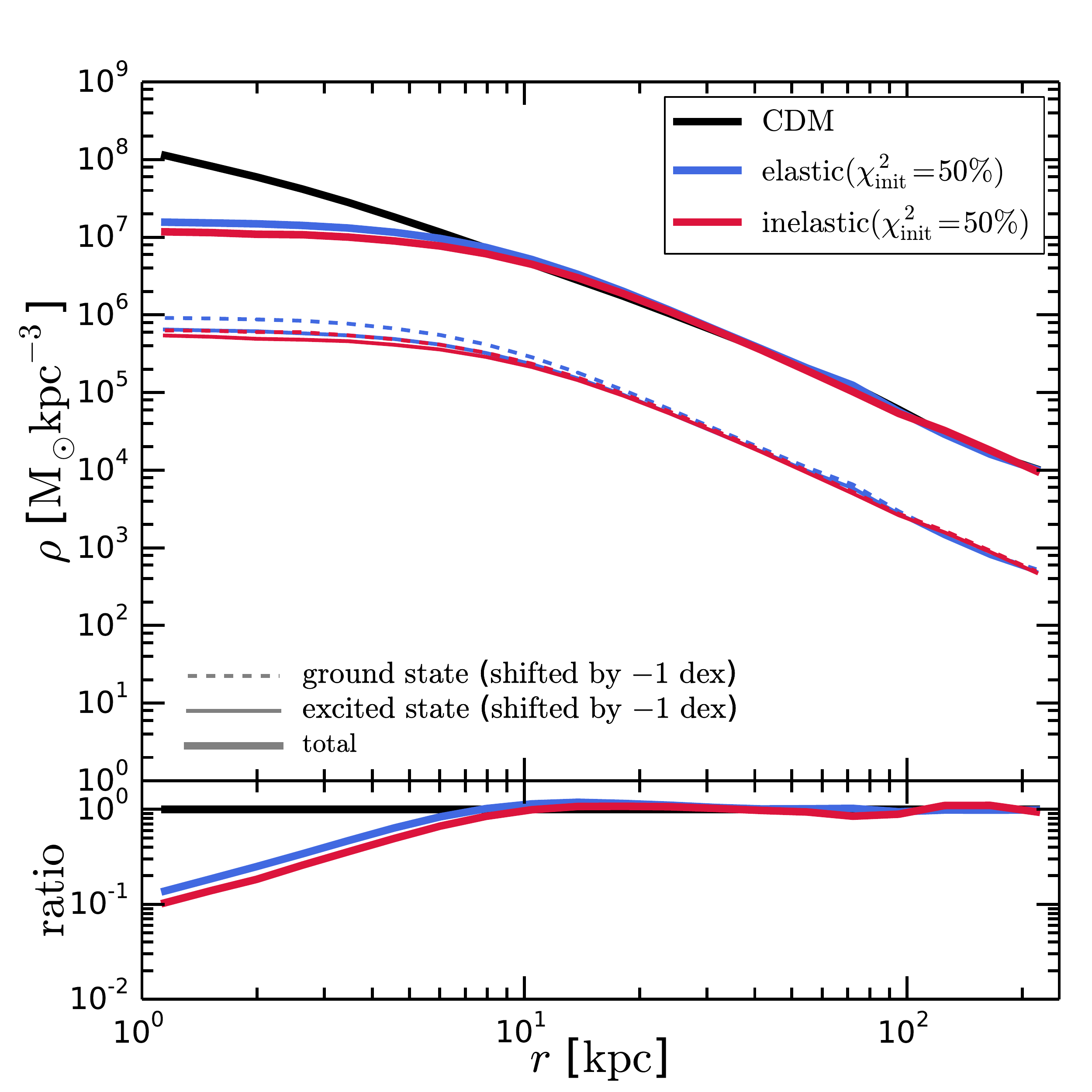}\\
\caption{{\bf Radial density profiles for the Milky Way-sized halo split in
the two-state populations.}  The inelastic SIDM models lead to larger cores with lower
central densities compared to the elastic SIDM models. This is caused by the
removal of ground state particles after exothermic reactions, which is largest
for the $\ICfull$ case, where more energy is available for release.  For the
inelastic case the total density almost never exceeds the CDM density at
intermediate radii.  This is distinct from the typical behaviour of elastic
SIDM models, where a small density enhancement over the CDM density is always observed due
to the redistribution of particles from the inner to the intermediate regions.
The ground state density profile for the $\ICfull$ configuration is very
different for the elastic and inelastic cases due to the removal of de-excited
particles from the halo centre. For the $\IChalf$ configuration, on the other
hand, the profiles are very similar since the de-excited particles represent
only a small fraction of the total ground state population. The bottom panels
show the ratio of the total density profiles relative to CDM.} 
\label{fig:rho_profiles}
\end{figure*}

The lower panels of Fig.~\ref{fig:frac_profiles} show the mass-weighted average
number of scattering events per particle in each reaction channel.  For  the
$\IChalf$ configuration (right panel) we find that the average number of
scattering events follows closely the overall ranking of the cross sections
presented in the bottom panel of Fig.~\ref{fig:schematic}.  For example, the mixed elastic reaction
(${\chi^1 + \chi^2 \rightarrow \chi^1 + \chi^2}$) has the largest cross section
of all channels and, correspondingly, it also contributes to the largest number
of average scattering events at all radii.  The exothermic reaction (${\chi^2 +
\chi^2 \rightarrow \chi^1 + \chi^1}$) on the other hand, has the smallest cross
section, except for very small relative velocities, and therefore leads to the
smallest number of scattering events. The inelastic model shows a lower number of
scattering events along that channel (long-dashed red lines) compared to the
elastic case (long-dashed blue lines), because of the removal of particles once
they de-excite into the ground state. The situation is quite different for the
$\ICfull$ configuration (left panel) since some channels are initially
suppressed because no ground state scattering partners can be found (e.g.
${\chi^1 + \chi^2 \rightarrow \chi^1 + \chi^2}$).  Also, given that all
particles are initially in the excited state, the number of scattering events for
the ${\chi^2 + \chi^2 \rightarrow \chi^2 + \chi^2}$ channel is the highest
despite having a smaller cross section than the mixed channel.  The thick solid
lines show the total number of scatters summed over all channels. These are
nearly the same for the elastic and inelastic cases in the $\IChalf$
configuration, whereas there is a clear difference in the $\ICfull$ case.
Again, this is caused by the removal of ground state particles following
de-excitation.  We note that for all cases, on average, only a few scatters per
particle occur in a Hubble time within the inner halo. This number is nearly
constant within the central $\sim20\kpc$ of the halo. Beyond $\sim 100\kpc$,
the number of scattering events drops below one rapidly.

\begin{figure*}
\centering
\hspace{-0.075cm}\includegraphics[width=0.495\textwidth]{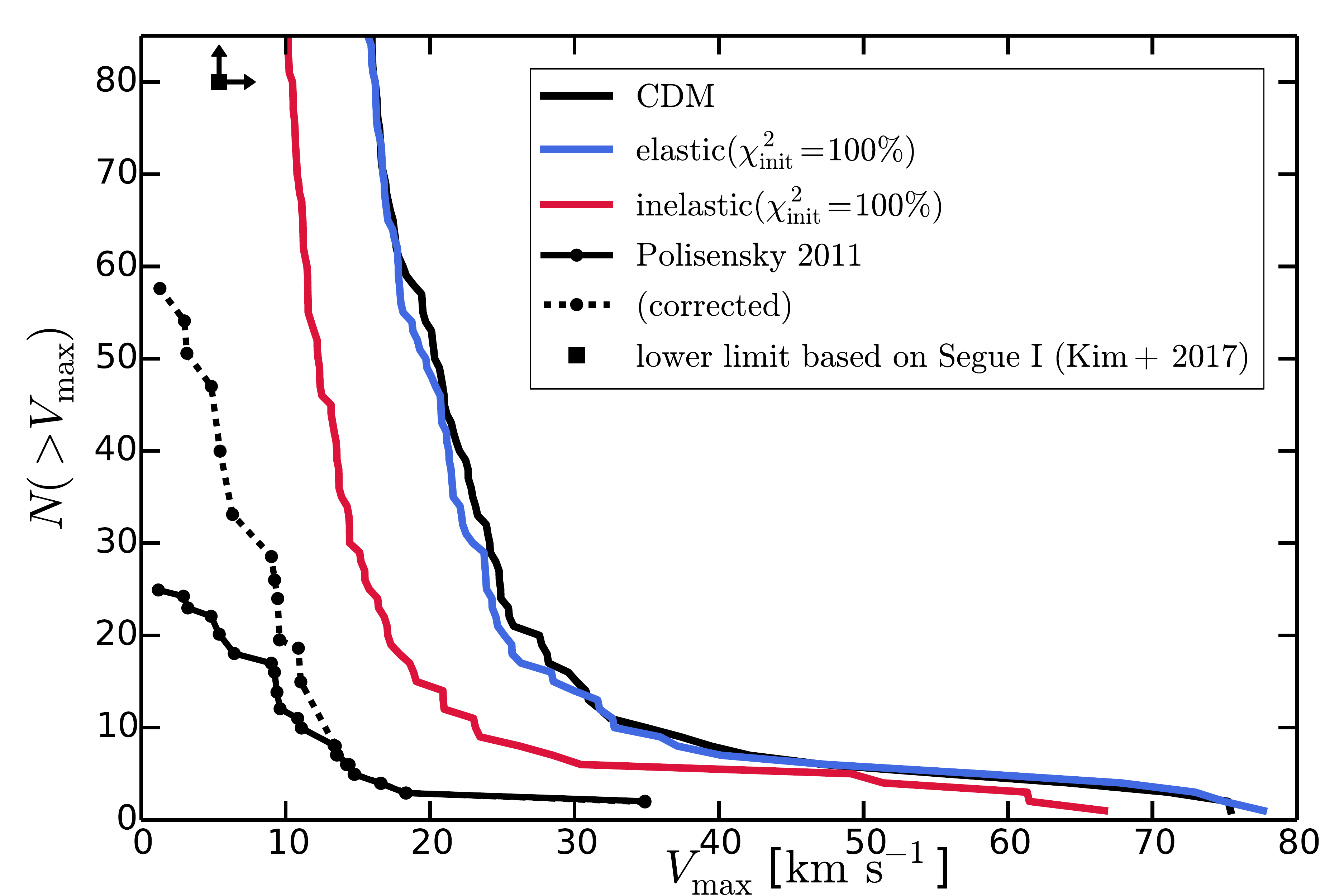}
\includegraphics[width=0.495\textwidth]{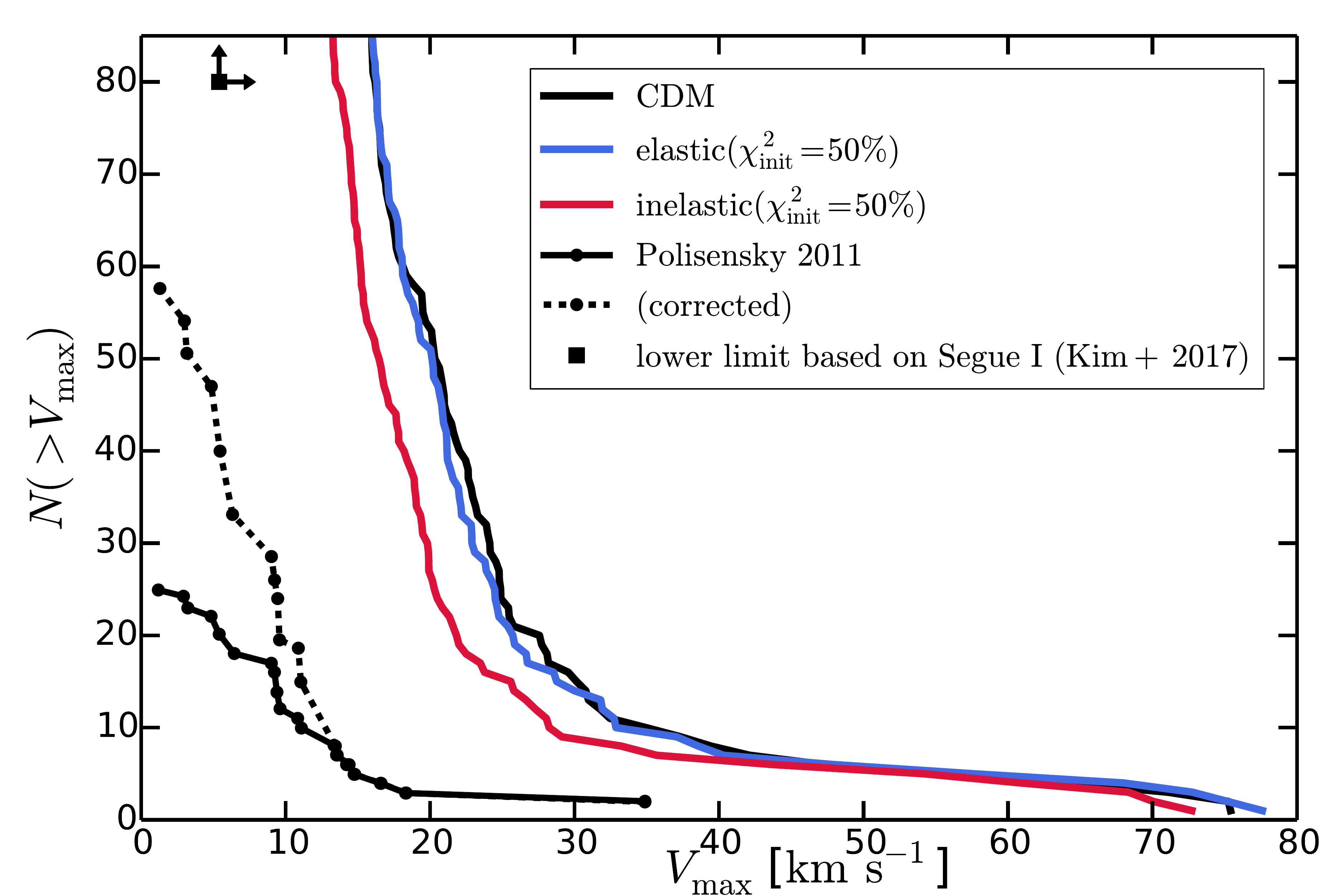}\\
\includegraphics[width=0.495\textwidth]{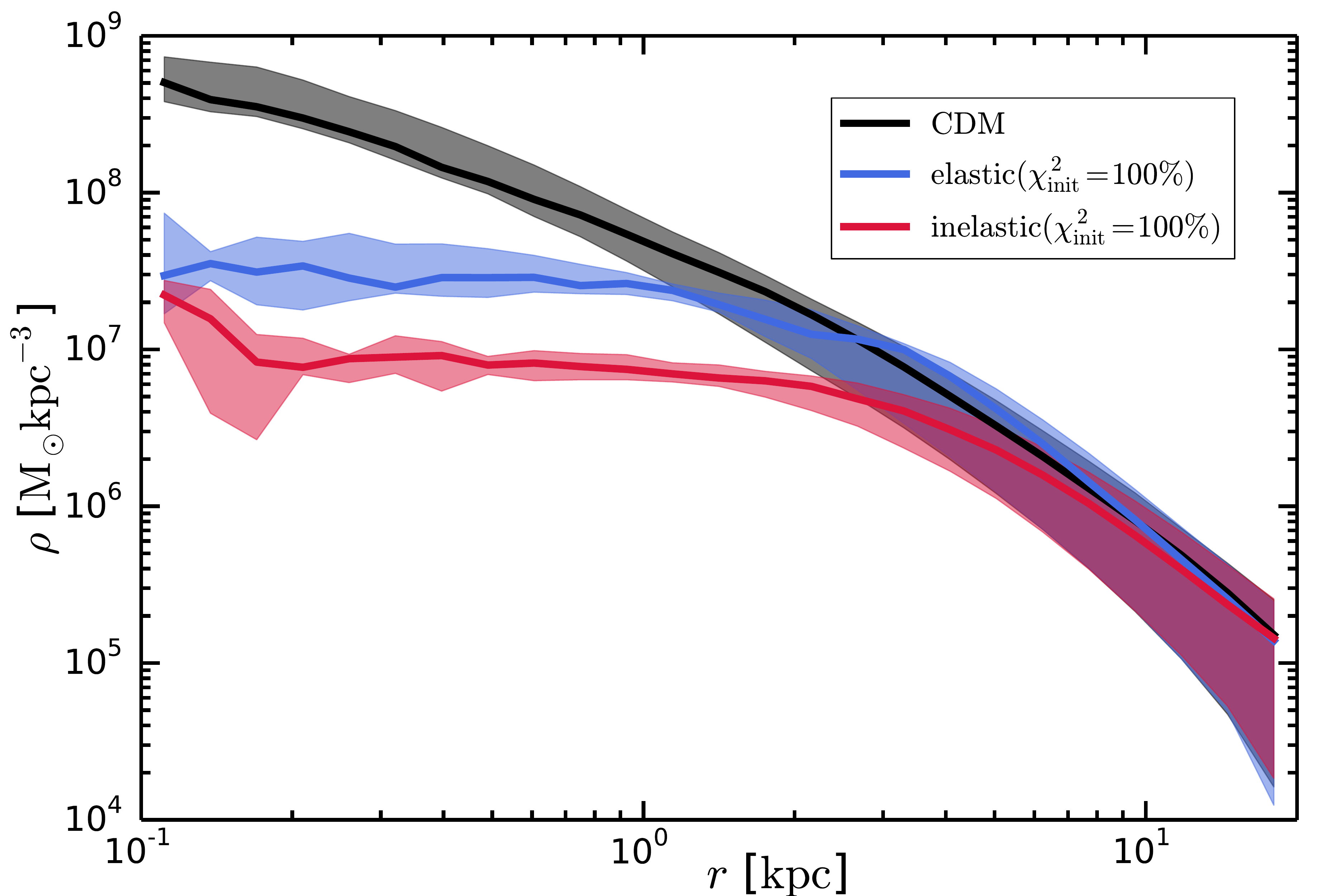}
\includegraphics[width=0.495\textwidth]{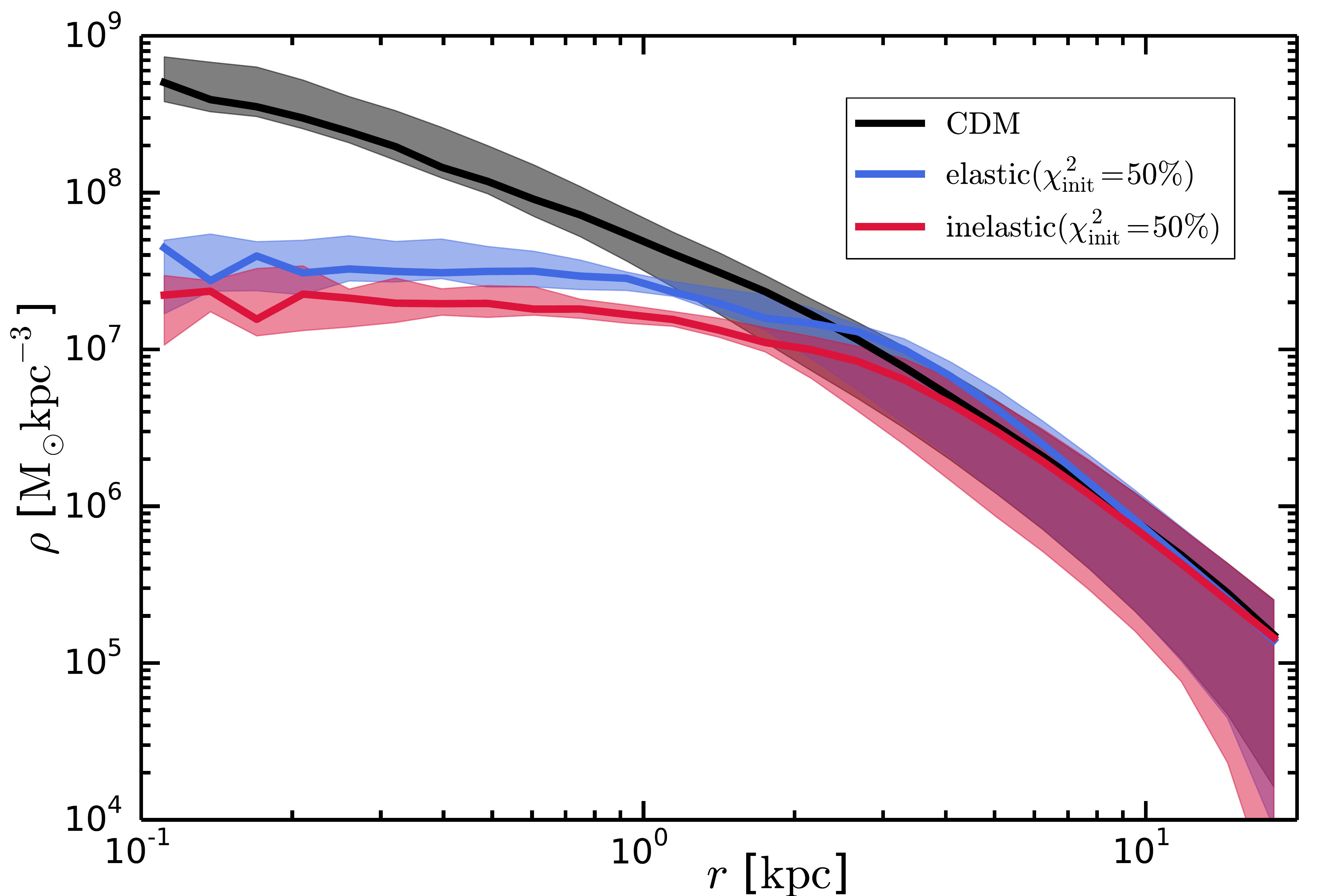}
\caption{{\bf Abundance and inner structure of subhaloes in CDM and (in)elastic
SIDM models.} {\it Top panels:} Cumulative subhalo velocity ($V_{\rm max}$)
function for the different models and observed satellites of the Milky
Way including a sky coverage correction~\citep[][]{Polisensky2011}.  Despite the relatively large cross sections (see
bottom panel of Fig.~\ref{fig:schematic}), the elastic model leads only to a minor reduction in
subhalo abundance relative to CDM.  A substantially  larger effect is visible
for the inelastic case.  For ${V_{\rm max} \gtrsim 30\kms}$ (${V_{\rm max}
\gtrsim 20\kms}$) the number of subhaloes is reduced by ${\sim 3}$ (${\sim 4}$)
for the inelastic model with $\ICfull$.  {\it Bottom panels:} Subhalo density
profiles. The solid lines show the median profile of the ten most massive
subhaloes at $z=0$, while the shaded region indicates the $1\sigma$ scatter of
the distribution. Inelastic SIDM leads to significantly larger and lower
density cores compared to the elastic case.  For the $\ICfull$ configuration
the core density is reduced by $\sim 4$ compared to the elastic case. This
implies that previous estimates on cross
sections within elastic SIDM simulations based on certain requirements for core densities, are altered in the
presence of inelastic SIDM reactions. We also note that such an inelastic SIDM
model will not suffer from the gravothermal catastrophe if the elastic
scattering cross sections are sufficiently small. }
\label{fig:subhalos}
\end{figure*}

\begin{figure*}
\centering
\includegraphics[width=0.495\textwidth]{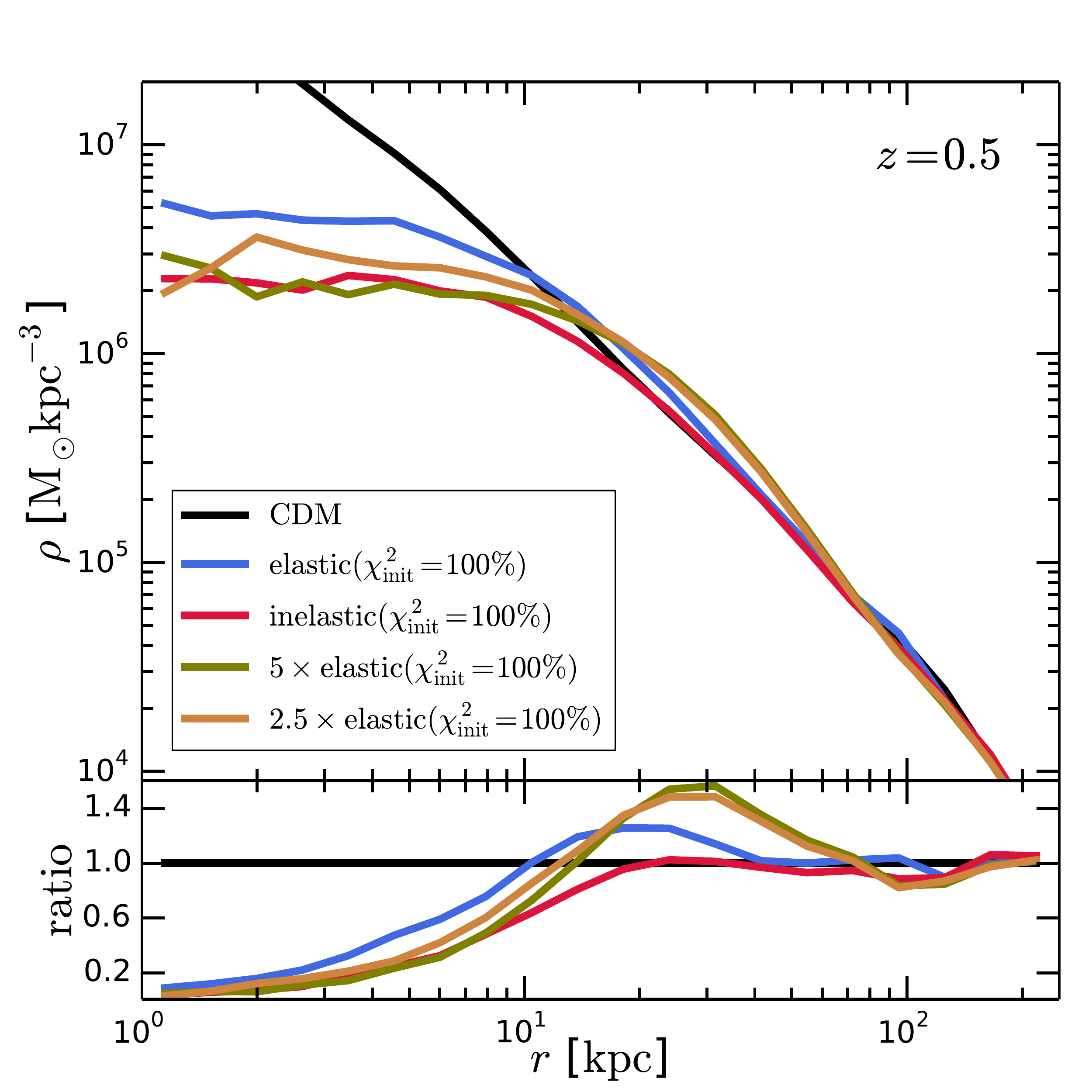}
\includegraphics[width=0.495\textwidth]{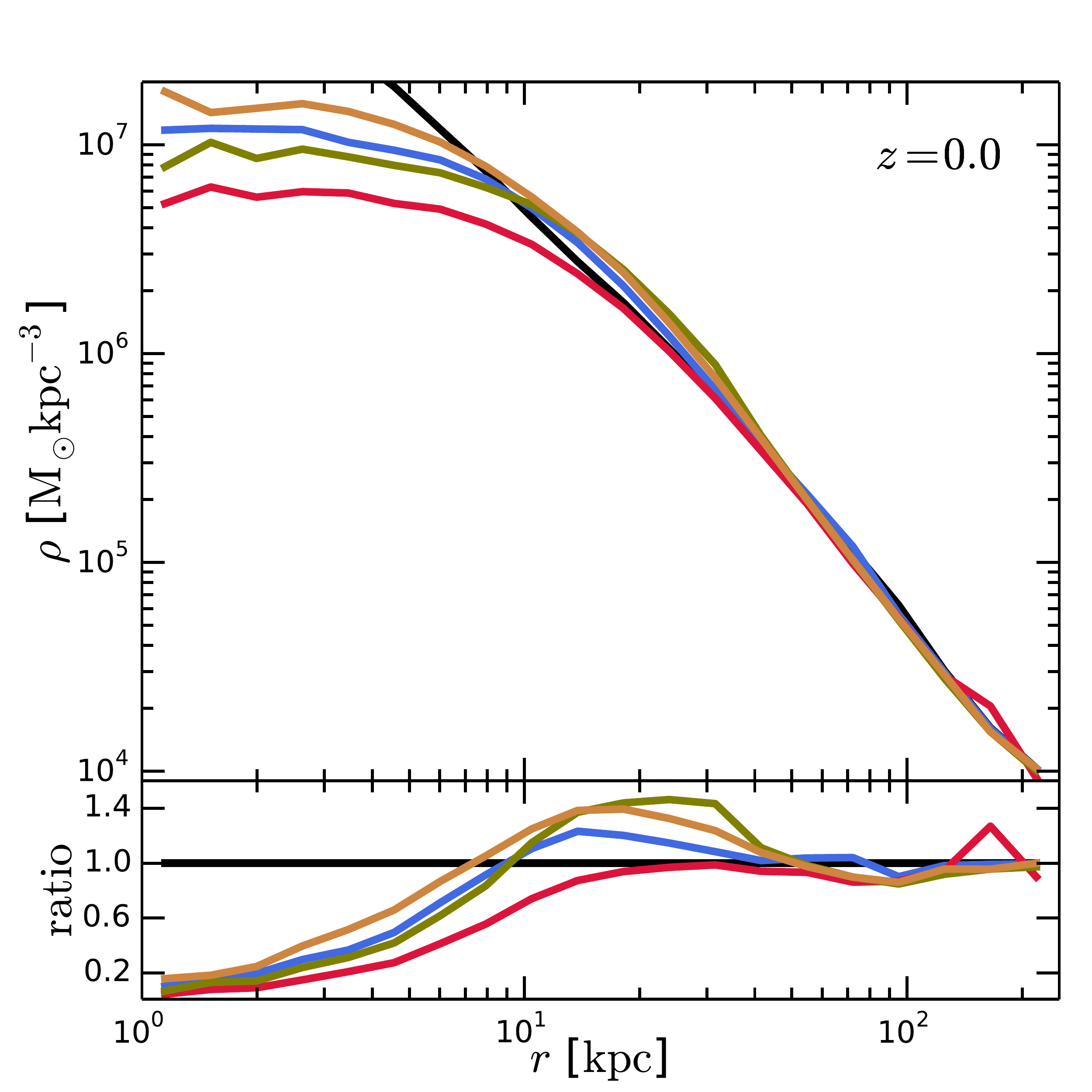}
\caption{{\bf Radial density profiles for the Milky Way-sized halo for scaled elastic cross section models.} An elastic model
with a $5$ times larger cross section leads to a central density reduction similar to the inelastic model at $z=0.5$ ({\it left panel}). The scaled elastic models core collapse at later times, but not the inelastic model ({\it right panel}). Inelastic models are therefore more efficient in core formation, and furthermore avoid the core collapse phase of elastic models.
This implies that current constraints on the normalisation of elastic cross sections need to be revised
if inelastic scatterings are the dominant mode of self-interactions. These simulations were performed at level-3 resolution, which is sufficiently converged (see Appendix~A).}
\label{fig:profile_rho_large_cross}
\end{figure*}

The net impact of energy release due to de-excitations is seen in
Fig.~\ref{fig:rho_profiles} where we show the spherically averaged total radial
density profiles (thick lines) for the CDM (black), elastic (blue) and
inelastic (red) SIDM models. Velocity dispersion profiles are presented in Appendix~B. As in previous figures, on the left (right) we
show the $\ICfull$ ($\IChalf$) case.  One can immediately see that the
inelastic case leads to a stronger depletion of the central density (a larger
density core) than the elastic case.  We stress that the elastic and inelastic
cases have exactly the same reaction channels and cross sections, with the only
difference being the energy release during down-scattering in the inelastic
case.  An interesting implication of this result is that it is possible in
inelastic models to create a core of the same size and density as in the
elastic case but with a smaller scattering cross section. 
This is an important point since it indicates that in the inelastic case, a wider range of cross section normalisations could significantly modify the inner dark matter density while remaining consistent with other constraints, compared with the range preferred by elastic SIDM simulations.

The thin lines in Fig.~\ref{fig:rho_profiles} show the density profiles of the
individual states, ground state ($\chi^1$) and excited state ($\chi^2$) with
solid and dashed lines, respectively. For clarity, we have shifted those
profiles down by one dex, relative to the total profiles. As anticipated, the
density profile for the ground state in the inelastic $\ICfull$ case (left
panel, dashed red line) is strongly suppressed.  The average density of the
ground state is more than two orders of magnitude lower than in the elastic
case, where particles are not kicked out from the halo.  This is not the case
for the $\IChalf$ configuration, where the profiles for the different state
populations are rather similar and nearly the same between the elastic and
inelastic cases.

The lower panels in Fig.~\ref{fig:rho_profiles} present the actual density
reduction compared to the CDM case for the four different SIDM scenarios.
Elastic collisions alone reduce the central density at $1\kpc$ already by an
order of magnitude compared to the CDM case.  As stated, the strongest
reduction is for the inelastic $\ICfull$ model where by $z=0$, the density is
lower by nearly a factor of $\sim 25$ at $1\kpc$.  Comparing the elastic-CDM
and inelastic-CDM ratio profiles in more detail, it becomes also clear that,
contrary to the elastic case, the inelastic case does not show an enhancement
over the CDM density at intermediate radii (for the $\ICfull$ case). This is
because in inelastic models the particles are not only redistributed within the
halo due to the effective inside-out heat transport caused by elastic
scattering, but they can, at least for our benchmark model, also be removed,
and  not contribute to the halo density profiles anymore.  This is a
distinctive signature between elastic and inelastic SIDM models since all
elastic models that produce a core also lead to such a density enhancement at
intermediate radii.  We finally note that the impact of inelastic
down-scattering in the halo density profiles strongly depends on the initial
level population of the excited state.  For the $50\%$ case (right panel of
Fig.~\ref{fig:rho_profiles}), we see only a rather small effect compared to the
elastic case. 

We expect that inelastic SIDM models have a stronger impact on the abundance of
subhaloes compared to purely elastic models. In fact, for elastic models a
quite large cross section on galactic scales ($\sim 10\cpm$) is needed to
create a relevant difference relative to CDM. We  present the subhalo velocity ($V_{\rm max}$) function for our MW-size simulations in the
top panels of Fig.~\ref{fig:subhalos}. Given the relatively large cross
sections of our benchmark model, we already see a mild impact on the abundance
of subhaloes for the elastic case.  However, the effect is much larger for the
inelastic case.  Specifically, we find that for ${V_{\rm max} \gtrsim 30\kms}$
(${V_{\rm max} \gtrsim 20\kms}$) the number of subhaloes is reduced by ${\sim
3}$ (${\sim 4}$) for the inelastic model with $\ICfull$. For the $\IChalf$
configuration we find a smaller effect, roughly reduced by a factor of 2.

In the bottom panel of Fig.~\ref{fig:subhalos} we show the median (solid lines) and 
$1\sigma$ region (shaded areas) of the distribution of density profiles of the ten 
most massive subhaloes at $z=0$. Similarly to the main halo density
profile, we find that inelastic SIDM models lead to subhaloes with larger cores and
smaller densities compared to the elastic case.  We note however, that the effect
of inelastic collisions is larger for these smaller systems with lower
velocity dispersion than the main MW halo 
since the cross section for down scattering strongly increases
towards lower relative velocities as shown in the bottom panel of Fig.~\ref{fig:schematic}, where the
cross section exceeds the equal state elastic cross sections for relative
velocities below $\sim 10\kms$.  

For the $\ICfull$ ($\IChalf$) configuration we find that the core density is
reduced by ${\sim 4}$ $(2)$ compared to the elastic case. Therefore, the inelastic SIDM cross sections can be
smaller by a factor of a few compared to the elastic case while creating a
core of similar size and density. We quantify this reduction in
Fig.~\ref{fig:profile_rho_large_cross}, where we compare the elastic and
inelastic model for the $\ICfull$ initial configuration with an elastic model,
where we have increased the cross section normalisation by factors of $2.5$ and $5$. These simulations
were performed at level-3 resolution, which is sufficiently converged for this test (see Appendix~A). The
figure shows that the central density reduction for the main halo density
profile is comparable for the inelastic model and the elastic model with a five times larger cross section (left panel).
This implies that previous conclusions obtained from elastic SIDM simulations
on the requirements for core sizes and densities given a certain cross section
normalisation, are strongly altered in the presence of inelastic scattering. We
also note that inelastic SIDM models will not suffer from the gravothermal
catastrophe, as long as down-scattering and the resultant energy release is
not suppressed. This is demonstrated in the right panel, where at $z=0$ the scaled elastic models
already started to core-collapse, which does not occur for the inelastic model.
This absence of core collapse is distinctly different from elastic SIDM
models, where the runaway collapse of the core is unavoidable on a time scale
that depends on the normalisation of the cross section \citep{Koda2011}.  As for
the main halo density profile, we find that the inelastic model does not lead
to a density enhancement relative to the CDM case at intermediate radii. As we
have argued, this is due to the expulsion of de-excited ground state particles
from the halo. 

Finally, we remark that the formation of a core in the inelastic case is a
combination of the redistribution of energy following elastic scatterings and
loss of DM particles during inelastic down-scattering. The degree to which
effect is more important depends on the primordial fraction of particles in the
excited state, and the hierarchy of the velocity-dependent cross sections of
the different scattering channels. A substantial mass loss in the halo centre
is a distinct feature of inelastic models relative to elastic SIDM, and it
might be of particular relevance for constraining the (inelastic) energy
deposition. In elastic SIDM for instance, the mass redistribution in the halo
due to elastic scattering can be reversed if the galaxy within is compact
enough~\citep[e.g.,][]{Elbert2018} or if the gravothermal
collapse phase has been triggered. This reversal has been invoked to argue that
in elastic SIDM, not all haloes are expected to be cored, and in fact the
natural diversity of the inner dark matter densities reflected in the rotation
curves of dwarf galaxies~\citep[][]{Oman2015} is to be expected for elasitic SIDM
with ${\sigma_{\rm T}/m \sim 1\,{\rm cm}^2\,{\rm g}^{-1}}$~\citep[][]{Kamada2017}. This reversal of cored
profiles into dense profiles would have only a limited extent in the case of
inelastic SIDM since the central mass loss is irreversible and the gravothermal
collapse cannot be triggered. Thus, it will be quite relevant to explore the
interplay of inelastic SIDM and baryonic physics to check under which
conditions they can develop cuspy profiles, which is a promising avenue to
constrain inelastic SIDM.

\section{Conclusions}
We have performed high resolution simulations of a galactic halo
within an inelastic self-interacting dark matter model using a newly developed
self-interaction dark matter implementation in the {\sc Arepo} code to study
multi-state dark matter models. We simulate a generic inelastic model of a
nearly degenerate two-state dark matter particle with an energy level splitting
of $10^{-6}$ with respect to its ground state, and an interaction cross section
of a few ${\rm cm^2}{\rm g}^{-1}$ on the scale of dwarf galaxies.

We demonstrate that the physics of inelastic self-interacting dark matter leads
to interesting new effects, which are distinctly different or absent in elastic
self-interacting dark matter models. Most importantly, dark matter particles
can be removed from potential wells, which causes large core formation with
lower densities compared to elastic models for the same interaction cross
section. Furthermore, inelastic collisions also lead to subhalo evaporation reducing efficiently
the abundance of subhaloes. Our main findings are:\\[-0.4cm]
\begin{itemize} 
\item Inelastic SIDM creates larger and lower density cores compared to elastic
models for the same cross section. Specifically, a $\sim 5$ times larger
elastic cross section would be required to achieve the same density reduction
and core formation as the inelastic model. The exact factor depends on the underlying inelastic dark matter model. This implies that currently
existing constraints on self-interaction cross sections have to be revised if
inelastic reactions are taken into account.
\item Similar to models with a power spectrum cutoff, we find that inelastic
SIDM can reduce the amount of substructure for rather small interaction cross
sections. The reduction of the subhalo abundance is due to gravitational
particle unbinding such that for ${V_{\rm max} \gtrsim 30\kms}$ (${V_{\rm max}
\gtrsim 20\kms}$) the number of subhaloes is reduced by ${\sim 3}$ (${\sim 4}$)
compared to the cold dark matter case.  
\item The de-excitation of the upper level to the ground state can inject the
energy equivalent of $\mathcal{O}(100)$ million Type II supernovae in Milky Way-like haloes. 
\item The gravothermal catastrophe that can occur in elastic SIDM models, can
be avoided in inelastic SIDM models. This is possible because of the unbinding of particles that de-excite during the scattering process. If this process
dominates over elastic scattering, a central density increase is not possible
anymore since the dark matter mass is efficiently removed.
\item The virial mass of a Milky Way-like halo is reduced by about $10\%$ compared to the cold dark matter
case for inelastic self-interacting dark matter. 
\item Density profiles for elastic and inelastic SIDM models are distinctly different. Elastic models
lead to an increased dark matter density at intermediate radii caused by particles being transferred
from the core region to outer parts. This effect is not present for inelastic models that can unbind particles.
For such models the density profile follows the CDM prediction outside of the core region.
\end{itemize}

We conclude that inelastic SIDM models have some properties, like the removal
of substructure and enhanced core formation, that are absent in typically
studied elastic SIDM models. Those features can have a significant impact on
properties of Milky Way-like dark matter haloes and can address many of the
small-scale challenges of the cold dark matter paradigm. Furthermore,
inelastic SIDM models are a rather generic feature of many dark matter particle
physics models beyond the canonical WIMPs, which have still not been detected.
It is therefore high time to consider also alternative dark matter particles,
like inelastic SIDM, and understand how structure formation progresses in those
models.  In the future it will be interesting to study these inelastic models
with baryonic galaxy formation
models~\citep[e.g.][]{Vogelsberger2013b,Vogelsberger2014b} to understand their
impact on galaxy formation in more detail. Furthermore, we also expect changes
in the detailed dark matter phase-space
structure~\citep[e.g.][]{Vogelsberger2009, Vogelsberger2013a} due to the energy
injection and particle removal in inelastic self-interacting models. 

\section*{Acknowledgements}

We thank Volker Springel for giving us access to {\sc Arepo}. 
MV acknowledges support through an MIT RSC award, a Kavli Research Investment Fund, NASA ATP grant NNX17AG29G, and NSF grants AST-1814053 and AST-1814259. JZ acknowledges support
by a Grant of Excellence from the Icelandic Research Fund (grant number
173929$-$051). TS is supported by the Office of High Energy Physics of the U.S. Department of Energy under grant Contract Numbers DE-SC00012567 and DE-SC0013999.
KS is supported by a Hertz Foundation Fellowship and a National
Science Foundation Graduate Research Fellowship. The simulations were performed
on the joint MIT-Harvard computing cluster supported by MKI and FAS.

\begin{appendix}

\section{Numerical Convergence}

\begin{figure}
\centering
\hspace{-0.2cm}\includegraphics[width=0.49\textwidth]{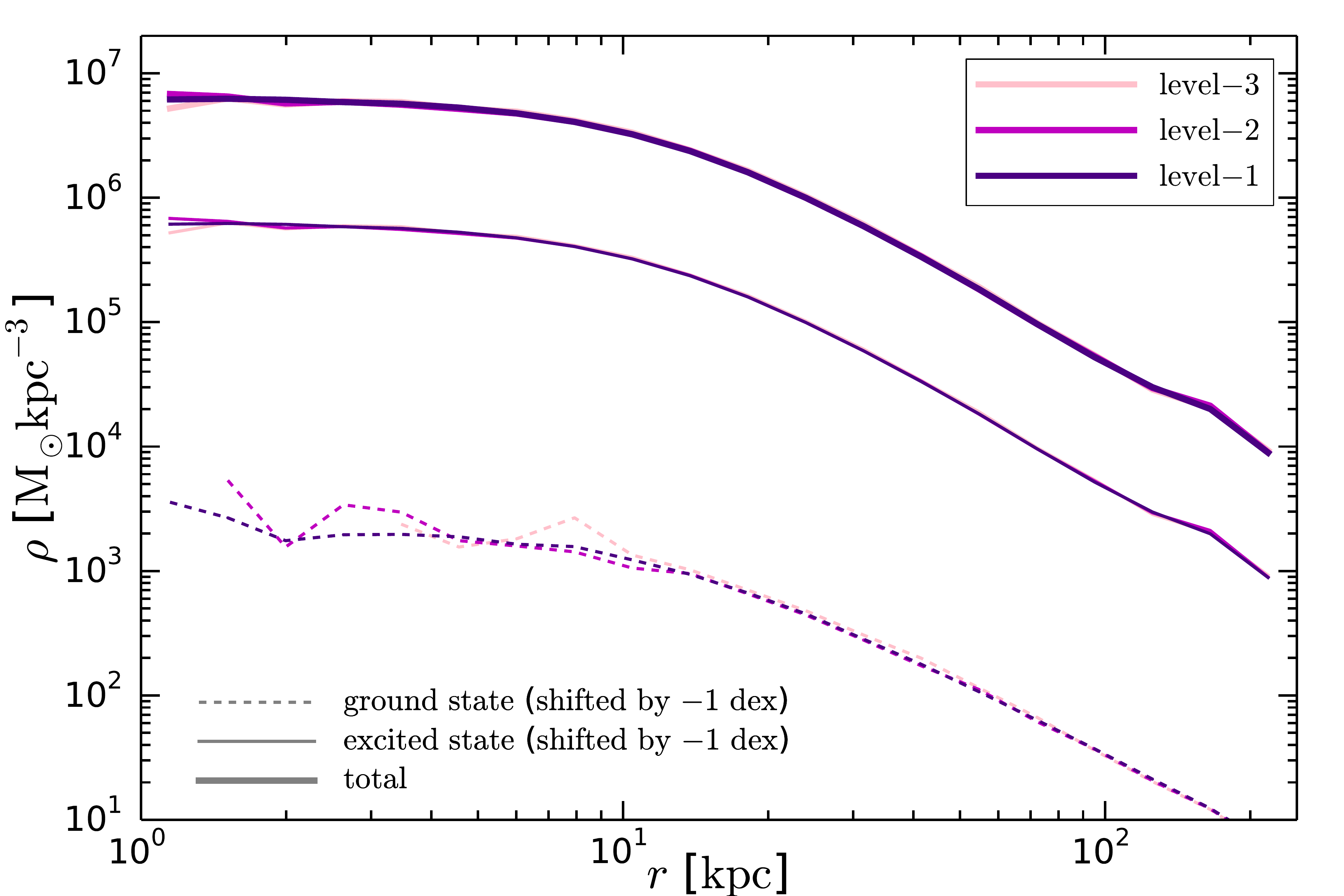}
\caption{{\bf Numerical convergence of inelastic SIDM simulations.} Density profiles for total density and state density split by ground and excited state for three different resolution levels of the $\ICfull$ initial configuration.}
\label{fig:convergence}
\end{figure}

We test the numerical convergence of the inelastic SIDM implementation in Figure~\ref{fig:convergence}. We present the density profiles for total density and state density split in ground and excited state for three different resolution levels of the $\ICfull$ initial configurations. The three resolution levels differ each by factors of $8$ in mass resolution, and a factor of $2$ in softening length. Level-1 corresponds to the highest resolution that has been employed for the paper. We find good convergence of the inelastic SIDM implementation. 

\section{Velocity Dispersion Profile}

\begin{figure}
\centering
\hspace{-0.2cm}\includegraphics[width=0.49\textwidth]{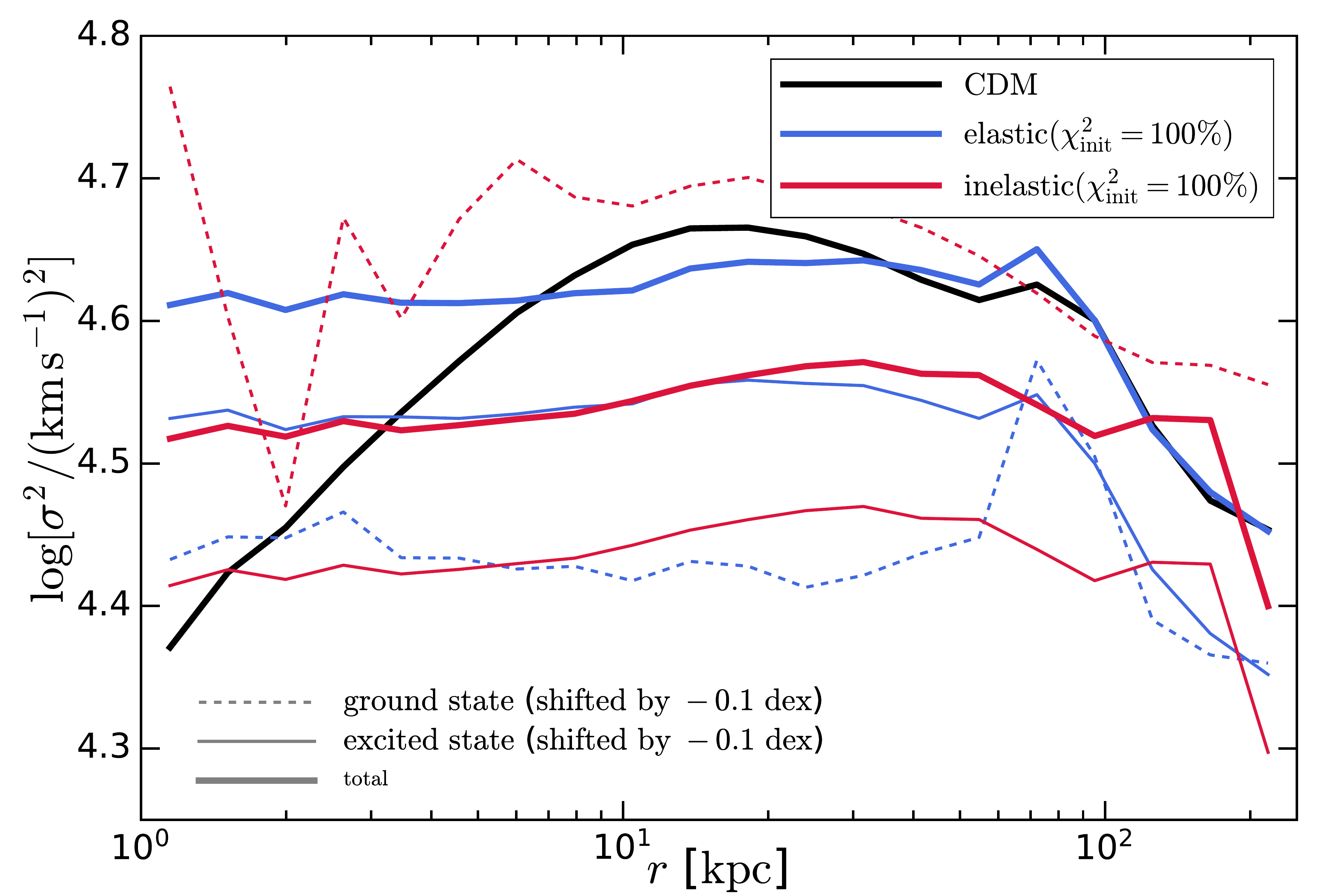}
\caption{{\bf Total velocity dispersion profiles for the Milky Way-sized halo split in the two-state populations.} Both elastic and inelastic SIDM lead to isothermal profiles in the centre. The ground state velocity dispersion is signficantly higher for the inelastic model due to the injected kinetic energy gained during the level decay.}
\label{fig:velocity_dispersion}
\end{figure}

The total velocity dispersion profile for the $\ICfull$ initial configuration is presented in Figure~\ref{fig:velocity_dispersion}. An isothermal core of very similar characteristics is formed in both the elastic and inelastic SIDM models, the key differences are that the inelastic case has a colder core (to compensate for the lower central density) and a population of ground state particles that is essentially unbound, moving with large velocities due to the energy injection during down-scattering.
\end{appendix}
\label{lastpage}

\end{document}